\documentclass[twocolumn]{aastex631}

\begin{document}
\title{\bf \large Signatures of the Small-scale Structure of the Pre-reionization Intergalactic Medium in $z\gtrsim7$ Quasar Proximity Zones}

\author[0000-0003-0821-3644]{Frederick B. Davies}
\affiliation{Max-Planck-Institut f\"{u}r Astronomie, K\"{o}nigstuhl 17, D-69117 Heidelberg, Germany}
\affiliation{Lawrence Berkeley National Laboratory, 1 Cyclotron Rd, Berkeley, CA 94720, USA}
\affiliation{Department of Physics, University of California, Santa Barbara, CA 93106-9530, USA}
\author[0000-0002-7054-4332]{Joseph F. Hennawi}
\affiliation{Department of Physics, University of California, Santa Barbara, CA 93106-9530, USA}
\affiliation{Leiden Observatory, Leiden University, NL-2300 RA Leiden, Netherlands}

\begin{abstract}
The small-scale structure of baryons in the intergalactic medium is intimately linked to their past thermal history. Prior to the $\gtrsim10^4$\,K photoheating during the epoch of reionization, cold baryons may have closely traced the clumpy cosmic web of dark matter down to scales as low as $\lesssim1$ comoving kpc, depending on the degree of heating by the X-ray background. 
After the passage of the ionization front, this clumpy structure can persist for $\sim10^{8}$ years.
The strong Ly$\alpha$ damping wings detected towards a few of the highest redshift quasars, in addition to their smaller-than-expected Ly$\alpha$-transmissive proximity zones, suggest that they have ionized and heated the foreground intergalactic medium less than $10^7$ years ago. Signatures of the pre-reionization small-scale structure should thus persist in their intergalactic surroundings. Here we explore how the persistence of this clumpy structure can affect the statistics of Ly$\alpha$ transmission inside the transparent proximity zones of $z\gtrsim7$ quasars by post-processing a suite of small-volume hydrodynamical simulations with 1D ionizing radiative transfer. We find that the Ly$\alpha$ flux power spectrum and flux PDF statistics of ten $z=7.5$ proximity zones, with realistic observational parameters, could distinguish the gaseous structure of a $T_{\rm IGM}\sim2$\,K CDM model from warm dark matter models with particle masses $m_{\rm WDM}>10$\,keV and X-ray heated models with $f_{\rm X}f_{\rm abs}>0.1$ ($T_{\rm IGM}(z=7.5)\gtrsim275$\,K) at the $2\sigma$ level.

\end{abstract}

\keywords{Intergalactic medium(813), Early universe(435), Reionization(1383)}

\section{Introduction}

The study of baryonic structure through the absorption by residual neutral hydrogen in the ionized intergalactic medium (IGM), the Ly$\alpha$ forest, is one of the foundations of modern observational cosmology. The Ly$\alpha$ forest represents the filaments and sheets of the cosmic web (e.g. \citealt{Meiksin09,McQuinn16}), and is the primary observable of cosmological structure formation in the linear and mildly non-linear regimes at higher redshifts ($z\sim2$--$5$) than is currently possible for large galaxy surveys.

Specifically, the Ly$\alpha$ forest enables measurements of the evolution of small-scale structure ($\lesssim100$\,comoving kpc) across a wide range of cosmic time (e.g. \citealt{Rorai13,Rorai17,Irsic17a,Walther18,Boera19}). These measurements have enabled some of the strongest cosmological constraints to date on the properties of dark matter \citep{Viel13,Irsic17b,Irsic17c,Irsic20,PalanqueDelabrouille20}. The Ly$\alpha$ forest is also sensitive to the thermal properties of the IGM, through the $\sim10^4$\,K Doppler broadening of absorption features and the $\sim100$~comoving kpc pressure-smoothing of the baryons relative to the dark matter (e.g. \citealt{GH98,Peeples10a,Rorai13,Kulkarni15}). The IGM thermal state is astrophysically interesting as well, as it traces the injection of energy by ionizing photons during the epochs of hydrogen and helium reionization, but the associated thermal broadening and pressure smoothing effects introduce confounding degeneracies to constraints on the underlying dark matter structure. 

\begin{figure*}[ht]
\begin{center}
\resizebox{15cm}{!}{\includegraphics[trim={3em 1.5em 1em 4em},clip]{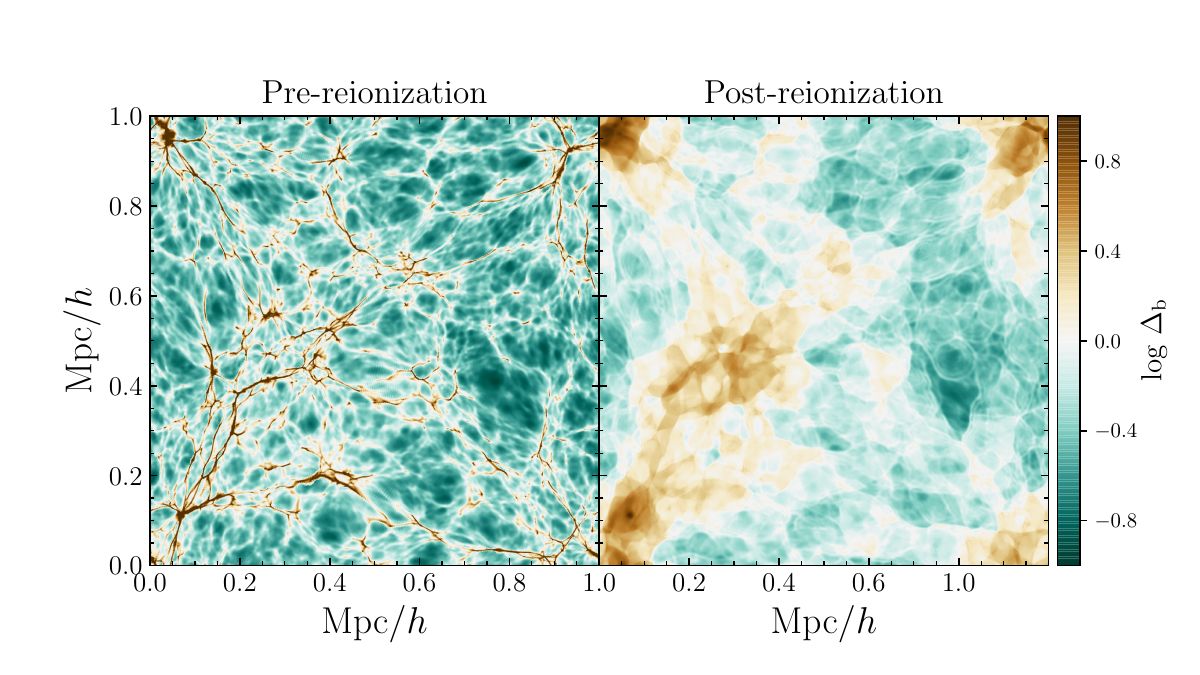}}\\
\end{center}
\caption{Slices through the baryon density field ($\sim1$ kpc$/h$-thick) of two of our hydrodynamical simulations at $z=7.5$. The gas in the simulation on the left has not yet been reionized, and thus is very cold ($\sim2$\,K), while the simulation on the right was reionized at $z=12$ by a uniform UV background and is thus much hotter ($\sim10^4$\,K) and significantly pressure-smoothed.}
\label{fig:hotcold}
\end{figure*}

Ideally, to minimize the effect of baryonic physics, one would like to probe the Ly$\alpha$ forest \emph{before} any reionization event. The thermal state of the IGM prior to hydrogen reionization is dramatically different from the typical post-reionization IGM probed by the Ly$\alpha$ forest, where adiabatic cooling can bring the IGM temperature down to as low as $T\sim2$\,K at $z\sim7$. Instead, however, it is widely accepted that the X-ray background built up by the radiation from accretion onto the first black holes and other compact objects (e.g., X-ray binaries) likely pre-heated the IGM to $T\sim100$--$1000$\,K prior to reionization, but the exact level is highly uncertain \citep{Furlanetto06Xray,Fialkov14,MF17,HERA22,HERA23}. This low gas temperature implies a much smaller pressure smoothing scale, i.e. the baryons should trace the dark matter down to scales as low as $\sim1$\,comoving kpc. To illustrate the structural difference between the pre- and post-reionization IGM, in Figure~\ref{fig:hotcold} we show slices of the baryon density field from two 1 Mpc$/h$ hydrodynamical simulations at $z=7.5$ (to be described in the next Section), one where reionization and its associated heat injection occurred much earlier at $z=12$ (left) and one where reionization has not yet occurred (right), highlighting the dramatic difference in baryonic structure caused by post-reionization pressure smoothing (see also \citealt{Park16,Hirata18,D'Aloisio20,Nasir21,Chan23}). Direct observation of the pre-reionization IGM at these epochs is impossible, unfortunately, due to the extreme opacity of the neutral IGM \citep{GP65}. The hydrodynamic response and resulting pressure smoothing of the gas occurs on $\gtrsim100$\,Myr timescales following the injection of heating by photoionization \citep{Park16,D'Aloisio20,Nasir21,Chan23}, so one potential solution would be to observe the Ly$\alpha$ forest \emph{soon after} gas is reionized at $z\sim6$--$8$. Unfortunately, the mean transmitted flux of the Ly$\alpha$ forest is already $\lesssim1\%$ at $z\sim6$ and decreases sharply at higher redshifts \citep{Eilers18,Bosman18,Yang20b,Bosman22}, greatly complicating any prospective precision analysis.

There is still one reionization-epoch environment whose Ly$\alpha$ forest can be probed at high precision: the Ly$\alpha$-transparent ``proximity zones'' of the most distant quasars. In the immediate intergalactic region around a quasar, the gas is subject to the quasar's strong ionizing radiation, leading to enhanced Ly$\alpha$ transmission within several proper Mpc along the line of sight \citep{Bajtlik88} which is especially pronounced at $z>6$ where the Ly$\alpha$ transmitted flux is otherwise close to zero (e.g. \citealt{Fan06,Eilers17,Davies20a,Satyavolu23}). The substantial transmission through the Ly$\alpha$ forest in quasar proximity zones has already enabled the current highest redshift ($z\sim6$) constraints on the IGM thermal state \citep{Bolton10,Bolton12}, and the ionizing background \citep{Calverley11}, but has not otherwise been explored for other detailed IGM analysis (although see \citealt{CG21b} for one possible avenue). The highest redshift quasars known at $z\sim7$--$7.6$ still exhibit Ly$\alpha$-transparent proximity zones \citep{Mortlock11,Banados18,Wang21}, despite appearing to lie deep within the reionization epoch (e.g.~\citealt{Greig17b,Greig19,Greig22,Davies18b,Wang20,Yang20a}), and their Ly$\alpha$ absorption profiles suggest they have likely ionized a substantial volume of the IGM along the line of sight for the first time \citep{Davies19}. Thus the proximity zones of the highest redshift quasars should contain gas exhibiting clumpy, pre-reionization signatures that is optimally sensitive to the nature of dark matter and early X-ray heating during cosmic dawn. The gaseous structures seen in Figure~\ref{fig:hotcold} are directly responsible for Ly$\alpha$ forest absorption, and the differences may be visible at high redshift provided they are illuminated by the intense ionizing radiation within a quasar proximity zone.

In this work, we show that the Ly$\alpha$ absorption structures in the proximity zones of $z\gtrsim7$ quasars can probe the baryonic structure of the pre-reionization IGM, and thus enable novel constraints on early X-ray heating and the small-scale structure of dark matter. We construct a simulation suite of small-volume high-resolution hydrodynamical simulations with varying degrees of small-scale structure, perform 1D radiative transfer of quasar ionizing radiation, and compute the resulting Ly$\alpha$ transmission spectra of proximity zones. Finally, we show that this effect can be statistically detected in the proximity zones of a realistic future sample of $z>7$ quasar spectra, providing novel insights into the clumpy (or not) nature of the pre-reionization IGM.

We assume $\Lambda$CDM and $\Lambda$WDM cosmologies with $h=0.675$, $\Omega_{\rm m}=0.315$, $\Omega_\Lambda=0.685$, and $\Omega_{\rm b}=0.049$, consistent with the most recent constraints from \emph{Planck} \citep{Planck18}. Distance units are comoving unless specified otherwise.

\section{Simulations of Quasar Proximity Zones in an Initially Neutral IGM}

To explore the effect of clumpy pre-reionization structure in the IGM on the proximity zones of $z\gtrsim7$ quasars, we run small-scale hydrodynamical simulations of the IGM, post-process them with 1D ionizing radiative transfer, and then compute the resulting Ly$\alpha$ transmission spectra. We describe these steps in detail below.

\subsection{Hydrodynamical Simulations}

\begin{table}
\begin{tabular}{l|c|c|c|c}
Name & $f_{\rm X}f_{\rm abs}$ & $m_{\rm DM}$\,(keV) & $z_{\rm re}$ & $T_{\rm IGM}$\,(K) \\
\hline \hline
Cold CDM & 0.0 & -- & $<7.5$ & 2.4 \\
Hot CDM & 0.0 & -- & $12$ & 10090 \\
X-ray0.01 & 0.01 & -- & $<7.5$ & 31 \\
X-ray0.1 & 0.1 & -- & $<7.5$ & 274 \\
X-ray0.3 & 0.3 & -- & $<7.5$ & 798 \\
X-ray0.9 & 0.9 & -- & $<7.5$ & 2352 \\
WDM5 & 0.0 & 5 & $<7.5$ & 1.7 \\
WDM10 & 0.0 & 10 & $<7.5$ & 1.7 \\
WDM20 & 0.0 & 20 & $<7.5$ & 1.9 \\
\end{tabular}
\caption{Hydrodynamical simulations in our simulation suite, each with 512$^3$ dark matter and baryon particles. The columns from left to right show the name of the simulation used in the text, the total number of (dark matter or baryon) particles, the product of the X-ray production efficiency and the fraction of energy absorbed by the IGM, the mass of warm dark matter particles (if any), the redshift of reionization, and the IGM gas temperature at mean density at $z=7.5$.}
\end{table}

We simulate the baryonic structure of the pre-reionization IGM with the smoothed-particle hydrodynamics (SPH) simulation code \texttt{MP-Gadget} \citep{MPGadgetDOI} with custom initial conditions to incorporate the initial relative velocity between baryons and dark matter \citep{TH10}. We also run simulations which include suppression of the primordial power spectrum by free-streaming of warm dark matter (WDM) \citep{Viel05} or add a redshift-dependent global heating of the IGM by X-rays \citep{Furlanetto06Xray} into the thermal evolution in the code. We discuss these modifications, as well as the parameters of our suite of simulations, below.

We use the code \texttt{CICASS} \citep{OM12} to generate cosmological initial conditions at $z=200$ for our \texttt{MP-Gadget} simulations. As discussed in e.g. \citet{Yoshida03} and \citet{OM12}, care must be taken to accurately simulate small-scale baryonic structures at early cosmic times. We adopt a similar simulation setup as \citet{OM12}, which we briefly summarize here. We generated separate baryon and dark matter spatial and velocity transfer functions at $z=1000$ with \texttt{CAMB} \citep{CAMB}, and evolved them following linear theory to $z=200$ with \texttt{CICASS}, taking into account the effect of a $30\,{\rm km}/{\rm s}$ relative velocity between baryons and dark matter ($v_{\rm bc}$; $\sim1\sigma$ fluctuation) on the evolution of the transfer function. We note, however, that at the redshift we are focused on in this work, $z=7.5$, the effect of $v_{\rm bc}$ is fairly small. 
The dark matter and baryon particles were initialized in a glass configuration with an initial offset of half the mean interparticle separation. Finally, linear perturbation theory was used to displace the particles to their initial positions at $z=200$ (see Appendix A of \citealt{OM12} for details).

\begin{figure}[ht]
\begin{center}
\resizebox{8cm}{!}{\includegraphics[trim={0em 1em 1em 1em},clip]{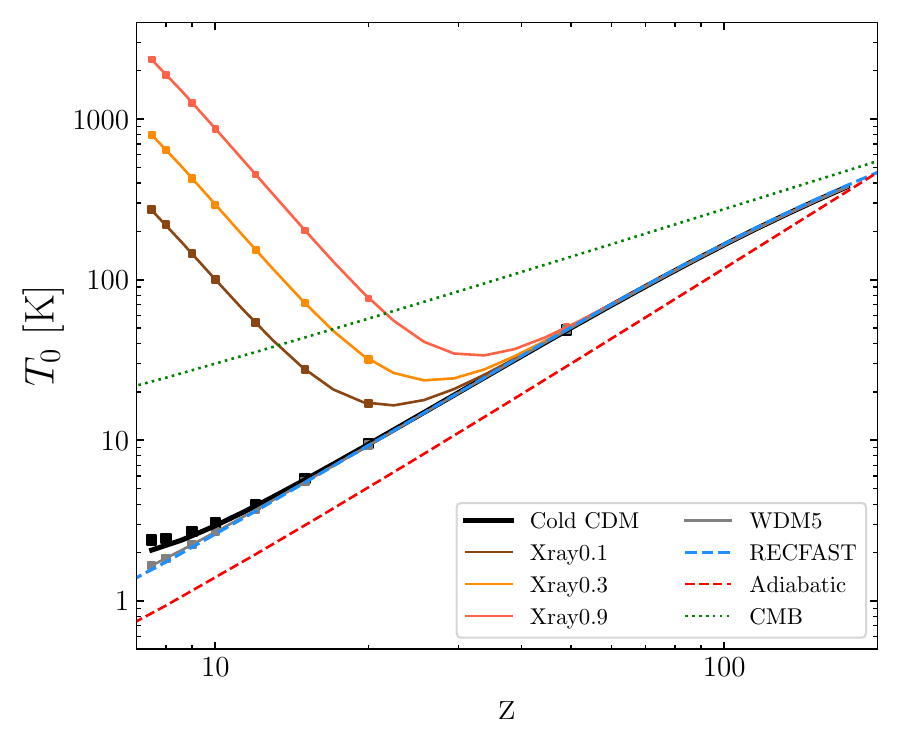}}\\
\end{center}
\caption{Evolution of the temperature at mean density in the Cold CDM (black), WDM5 (grey), X-ray0.1 (brown), X-ray0.3 (orange), and X-ray0.9 (red) simulations. Solid curves show finely spaced snapshots from $128^3$ simulations, while the squares show snapshots from our fiducial $512^3$ simulations. The evolution predicted by \texttt{RECFAST} is shown by the dashed blue curve, while simple adiabatic evolution due to the expansion of the Universe, $T_0\propto(1+z)^2$, and the CMB, $T_{\rm CMB}\propto(1+z)$, are shown by the dashed red and dotted green curves, respectively.}
\label{fig:tevol}
\end{figure}

For the WDM simulations, we modified the initial transfer functions using the analytic expression from \citep{Viel05}. As in many other works, we neglect the effect of relic thermal velocities due to their tendency to dramatically \emph{increase} power on small scales due to shot noise when implemented via applying random kicks to dark matter particles (e.g. \citealt{BD16,Leo17}), and note that these could further reduce the growth of structure on the small scales investigated here, albeit likely only modestly. 

\begin{figure*}[ht]
\begin{center}
\resizebox{17cm}{!}{\includegraphics[trim={3em 1.5em 2em 4em},clip]{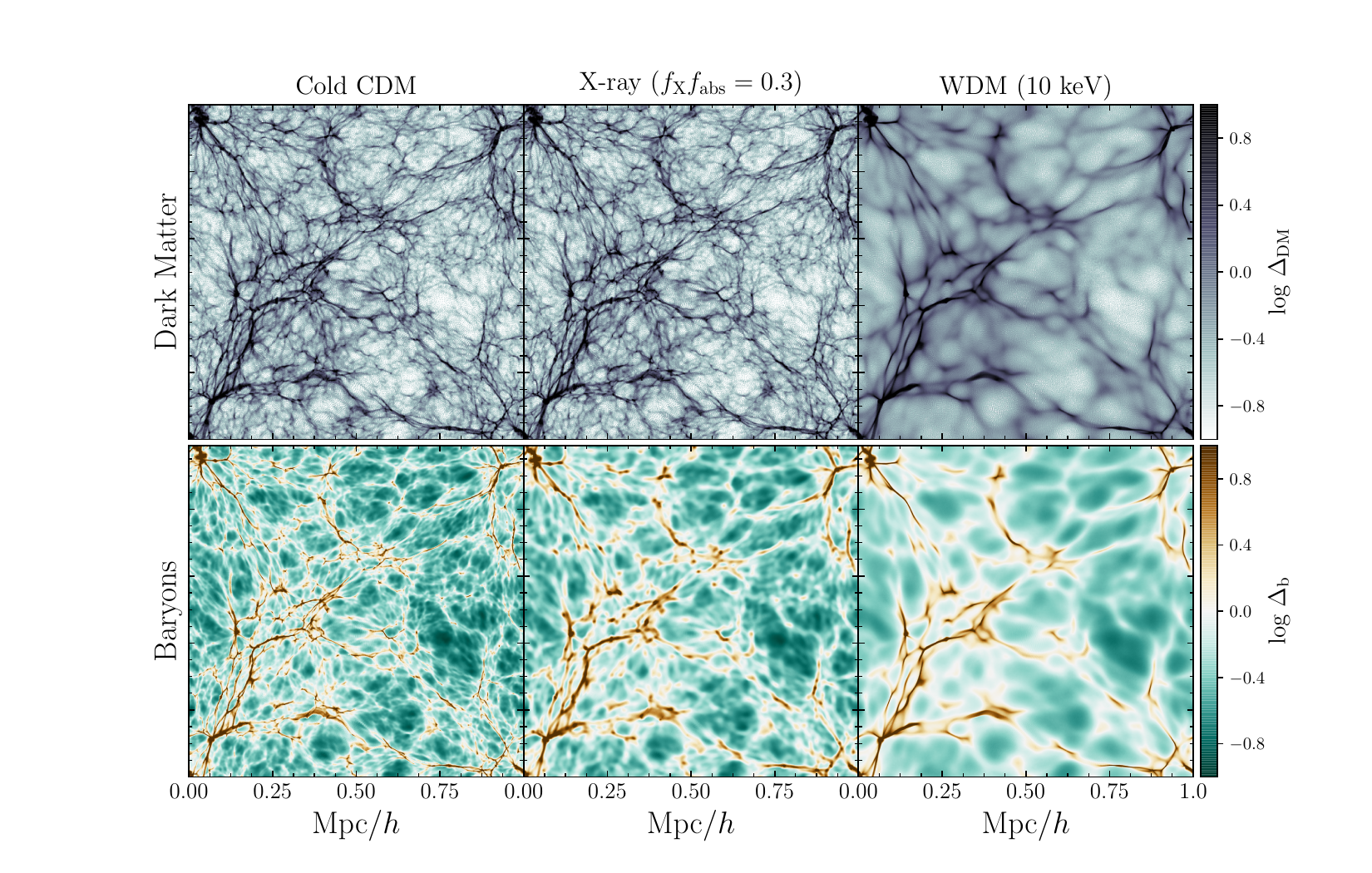}}\\
\end{center}
\caption{Slices through the dark matter (top) and baryon (bottom) density fields ($\sim4$ kpc$/h$-thick) of three of our hydrodynamical simulations at $z=7.5$. Left: Cold CDM model where the baryons have cooled roughly adiabatically since thermally decoupling from the CMB at $z\sim150$, resulting in gas temperatures of $T\sim2$\,K and a correspondingly tight connection between dark matter and baryon structure. Middle: X-ray heated model with $f_{\rm X}f_{\rm abs}=0.3$ (Xray0.3), corresponding to a total heat injection of $\sim800$\,K at this redshift. Compared to the left panels, the baryon distribution exhibits a modest amount of pressure-smoothing due to the elevated gas temperature. Right: Warm dark matter model with $m_{\rm WDM}=10$\,keV (WDM10), showing significantly suppressed small-scale structure in the dark matter which is also reflected in the baryons.}
\label{fig:slices}
\end{figure*}

The basic difficulty in simulating the pre-reionization small-scale structure is the required contrast in scales. We require $\sim1$\,comoving kpc resolution to resolve the Jeans scale of the cold gas, but the scales of $z\sim7$ quasar proximity zones are on the order of $1$--$2$\,proper Mpc \citep{Mortlock11,Banados18,Wang20,Yang20a}, also similar to the scale at which the fundamental mode of the simulation volume remains linear. We focus here on resolving the small scales, and run Lagrangian simulations (e.g. \citealt{Park16,Chan23}) instead of a uniform grid (e.g. \citealt{D'Aloisio20,Doughty23}) to resolve the moderately overdense gas responsible for Ly$\alpha$ forest absorption at the high transmission values probed inside of quasar proximity zones.

Our fiducial simulations were run with $512^3$ dark matter and baryon particles in boxes with side length $L=1\,{\rm Mpc}/h$ down to $z=7.5$, roughly corresponding to the redshift of the most distant quasars currently known \citep{Banados18,Yang20a,Wang21}. We adopt a gravitational smoothing length of $0.04$ times the mean interparticle spacing for the dark matter particles ($0.078$\,kpc) and use ``adaptive'' smoothing for the baryon particles (i.e., gravitational smoothing equal to the scale of the SPH smoothing kernel), following the suggestions in \citet{OM12}. For computational efficiency, we adopt the cubic spline SPH smoothing kernel (with $N_{\rm neigh}=44$ neighbors) instead of the \texttt{MP-Gadget} default quintic spline, and allow for the densest environments (with baryon overdensities $\rho_{b}/\bar{\rho_b}>1000$ and $T<10^5$\,K) to form collisionless star particles \citep{Viel04}.\footnote{We have verified that this truncation at the highest overdensities does not appreciably impact our results due to their very small volume filling factor.} 

In \texttt{MP-Gadget}, as in many hydrodynamical simulation codes, the electron fraction is computed assuming collisional and photoionization equilibrium by default. However, during the cosmic dark ages, the electron fraction remains substantially higher than its equilibrium value due to ``freeze-out'' after recombination at $z\sim1100$. This excess residual electron fraction is crucial to reproducing the expected temperature evolution, as Compton heating by the CMB continues to non-trivially heat the gas until at least $z\sim100$, resulting in temperatures a factor of $\sim2$ higher at $z\lesssim10$ compared to adiabatic cooling since the start of the simulations at $z=200$. To approximately recover the expected residual electron fraction in our simulations (similar to \citealt{Cain20}), we pre-compute the redshift-dependent freeze-out of electrons after recombination with \texttt{RECFAST} \citep{Seager99}, and use an interpolation table of $x_{\rm e}(z)$ to set a lower limit\footnote{We use a lower limit as structure formation shocks will collisionally ionize some regions to a higher electron fraction.} to the electron fraction in the thermal evolution calculated by \texttt{MP-Gadget}. With the correct $x_{\rm e}(z)$, we find that the temperature evolution at mean density predicted by \texttt{RECFAST} is accurately reproduced until the onset of shock heating at $z\lesssim10$--$20$, as shown in Figure~\ref{fig:tevol}.

For simulations with X-ray heating, we introduce an additional heating term to \texttt{MP-Gadget} following a procedure similar to \citet{Pober15}. We assume the empirical fit to the global star formation rate density
evolution from \citet{Robertson15} and associate this star formation with X-ray emission according to \citep{Furlanetto06Xray}:
\begin{equation}
L_{\rm X} = 3.4\times 10^{40} f_{\rm X} \left(\frac{{\rm SFR}}{1\ {\rm M}_\odot\,{\rm yr}^{-1}}\right)\ {\rm erg}\ {\rm s}^{-1},
\end{equation}
where $f_{\rm X}$ is a free parameter which represents our uncertainty in the efficiency of X-ray emission in the early Universe. The resulting heating rate is then
\begin{equation}
\epsilon_{\rm X} = 3.4\times10^{40} {\rm erg}\ {\rm s}^{-1}\ {\rm Mpc}^{-3}\ f_{\rm X} f_{\rm abs} \dot{\rho}_{\rm SFR}(z),
\end{equation}
where $f_{\rm abs}$ is the uncertain fraction of the energy in X-rays which contributes to the thermal energy of the IGM, which is related to their outgoing spectrum from galaxies and losses due to secondary photoionizations (e.g. \citealt{Fialkov14}), and $\dot{\rho}_{\rm SFR}$ is the star formation rate density in units of $M_\odot\ {\rm yr}^{-1}\ \rm{Mpc}^{-3}$. Given the substantial uncertainty in both $f_{\rm X}$ and $f_{\rm abs}$ we choose to parameterize our models in terms of the product $f_{\rm X}f_{\rm abs}$. We find that injecting the thermal energy from X-rays over time, rather than imposing a minimum temperature as in \citet{D'Aloisio20}, reduces their cumulative effect on the pressure smoothing of the baryons, but the difference is qualitatively small.

The suite of hydrodynamical simulations and their parameters are given in Table 1. 
We explore WDM masses of 5 to 20 keV, and X-ray heat injection of $f_{\rm X}f_{\rm abs}=0.01$ to $0.9$ resulting in global gas temperatures of $31$ to $2352$\,K, respectively, at $z=7.5$. In Figure~\ref{fig:slices} we show slices through the dark matter and baryon density fields of the Cold CDM, 10 keV WDM (WDM10), and $f_{\rm X}f_{\rm abs}=0.3$ X-ray heated (X-ray0.3) simulations. In the Cold CDM and WDM simulations, the baryons closely trace the dark matter, while in the X-ray simulation, there is clear pressure smoothing of the gas due to the additional thermal pressure.

\begin{figure}[ht]
\begin{center}
\resizebox{8.5cm}{!}{\includegraphics[trim={1em 1em 1em 1em},clip]{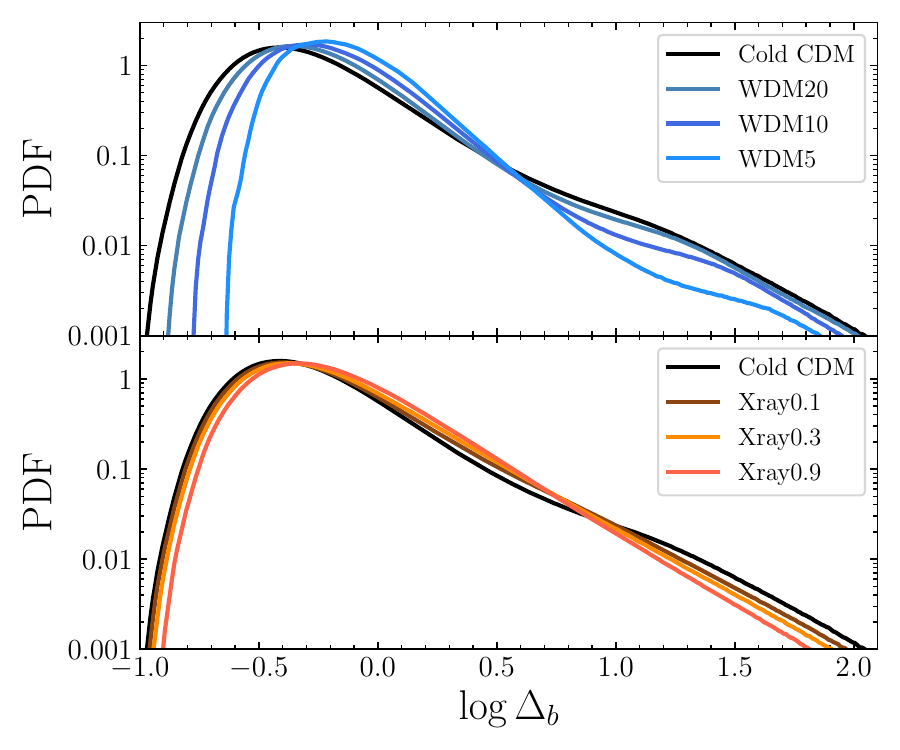}}\\
\end{center}
\caption{Distribution of baryon overdensities in the Cold CDM simulation (black curves) compared to the WDM models (upper panel) and X-ray heated models (lower panel).}
\label{fig:densitypdf}
\end{figure}

\begin{figure*}[ht]
\begin{center}
\resizebox{18cm}{!}{\includegraphics[trim={1em 2em 1em 1em},clip]{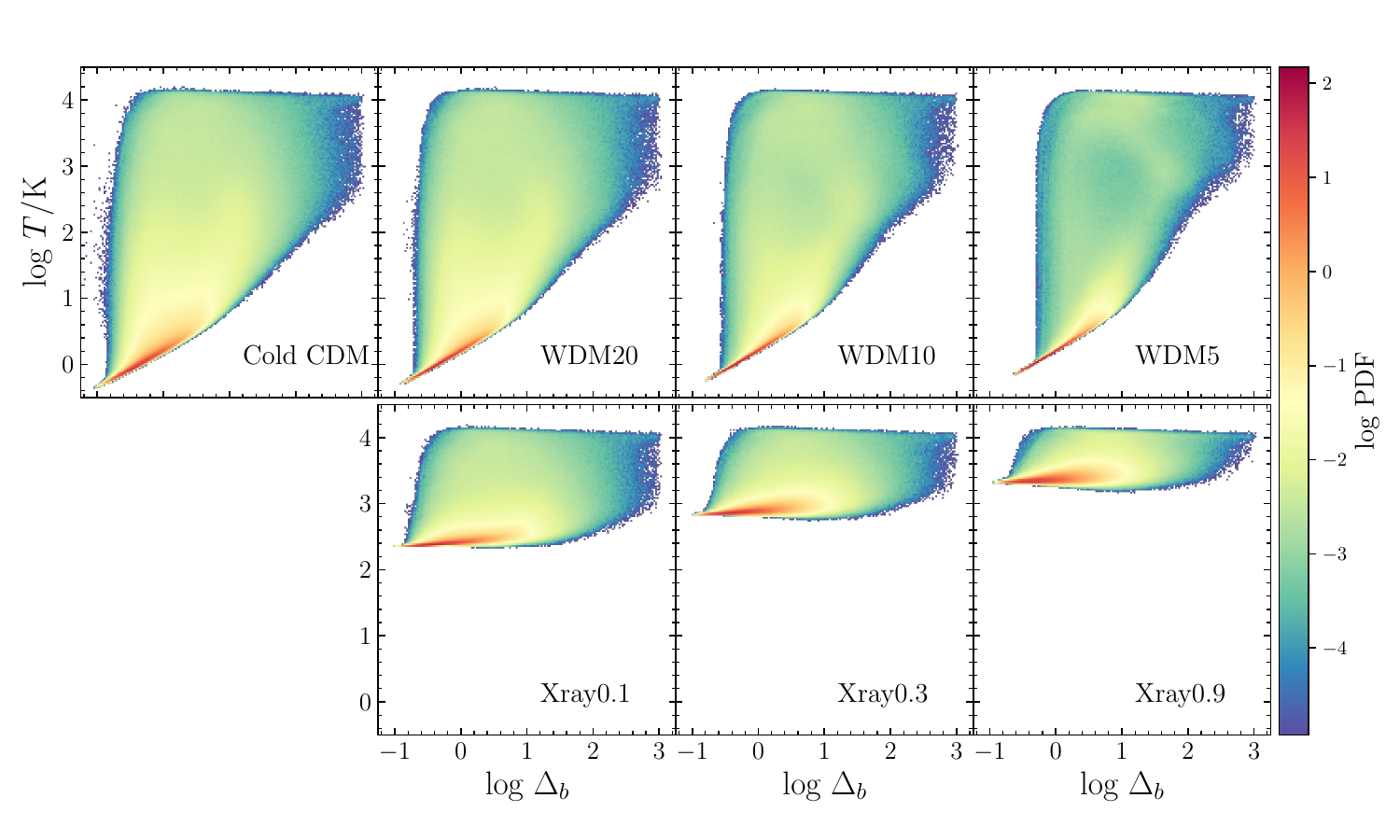}}\\
\end{center}
\caption{Relationship between baryon overdensity and gas temperature in the Cold CDM and WDM models (top), and the X-ray heated models (bottom). Heating by a uniform X-ray background introduces an effective temperature floor in the X-ray heated models compared to the much colder temperatures reached in the Cold CDM and WDM models.}
\label{fig:rhot}
\end{figure*}

We show the evolution of the temperature at mean density $T_0$ for several of our simulations in Figure~\ref{fig:tevol} compared to adiabatic cooling ($T_0\propto(1+z)^2$) and the \texttt{RECFAST} calculation. The solid curves show densely spaced outputs from $128^3$ simulations, capturing the full thermal evolution at higher temporal resolution, while the squares show outputs from our fiducial $512^3$ simulations. The IGM cools significantly more slowly than adiabatic at $z\gtrsim100$ due to the impact of inverse Compton heating by CMB photons on the residual electrons from recombination freeze-out. The thermal evolution of the X-ray models highlights the gradual heat injection at $z\lesssim20$, with the final $T_0$ at $z=7.5$ proportional to $f_{\rm X}f_{\rm abs}$. The Cold CDM model generally follows the \texttt{RECFAST} model, confirming that our $x_{\rm e}$ adjustment in \texttt{MP-Gadget} was sufficient to reproduce the correct thermal evolution, and shows a slight uptick in $T_0$ at $z\lesssim10$ due to shock heating which is slightly stronger in the higher resolution simulation (i.e. the $512^3$ black points vs. $128^3$ black curve). The relative absence of structure in the WDM5 model leads to somewhat weaker late time shock heating in comparison. 

The probability distribution functions (PDFs) of baryon overdensity in the simulations are shown in Figure~\ref{fig:densitypdf}, comparing the Cold CDM simulation to the WDM simulations in the upper panel, and to the X-ray heated simulations in the lower panel. As expected from the visual appearance of Figure~\ref{fig:slices}, lowering the WDM particle mass and increasing the X-ray heating both act to reduce the variance in the gas density, with corresponding reductions in the PDF at the upper and lower extrema.

The relationships between density and temperature in the simulations at $z=7.5$ are shown in Figure~\ref{fig:rhot}. While the majority of the gas in the Cold CDM and WDM simulations roughly falls along a power-law relationship between the gas temperature $T$ and the gas overdensity $\Delta$ established by the adiabatic expansion and contraction from quasi-linear structure formation, the low density gas in the X-ray heated simulations is much closer to isothermal due to the uniform heat injection by the X-ray heating prescription. 
The apparent $\sim10^4$\,K upper limit to the temperature in each simulation is set by the rapid onset of collisional ionization and its associated strong Ly$\alpha$ excitation cooling in regions shock-heated by structure formation. In the WDM simulations, the suppression of small-scale structure reduces the amount of shock heating, thus decreasing the total mass contained in the plume of material above the cold adiabatic locus. Note that in these simulations we have explicitly turned off the ionization and heating by the UV background, such that they remain relatively cold and neutral down to $z=7.5$.

\subsection{Ionizing Radiative Transfer}

\begin{figure*}
\begin{center}
\resizebox{17.5cm}{!}{\includegraphics[trim={1.0em 1em 1.0em 1em},clip]{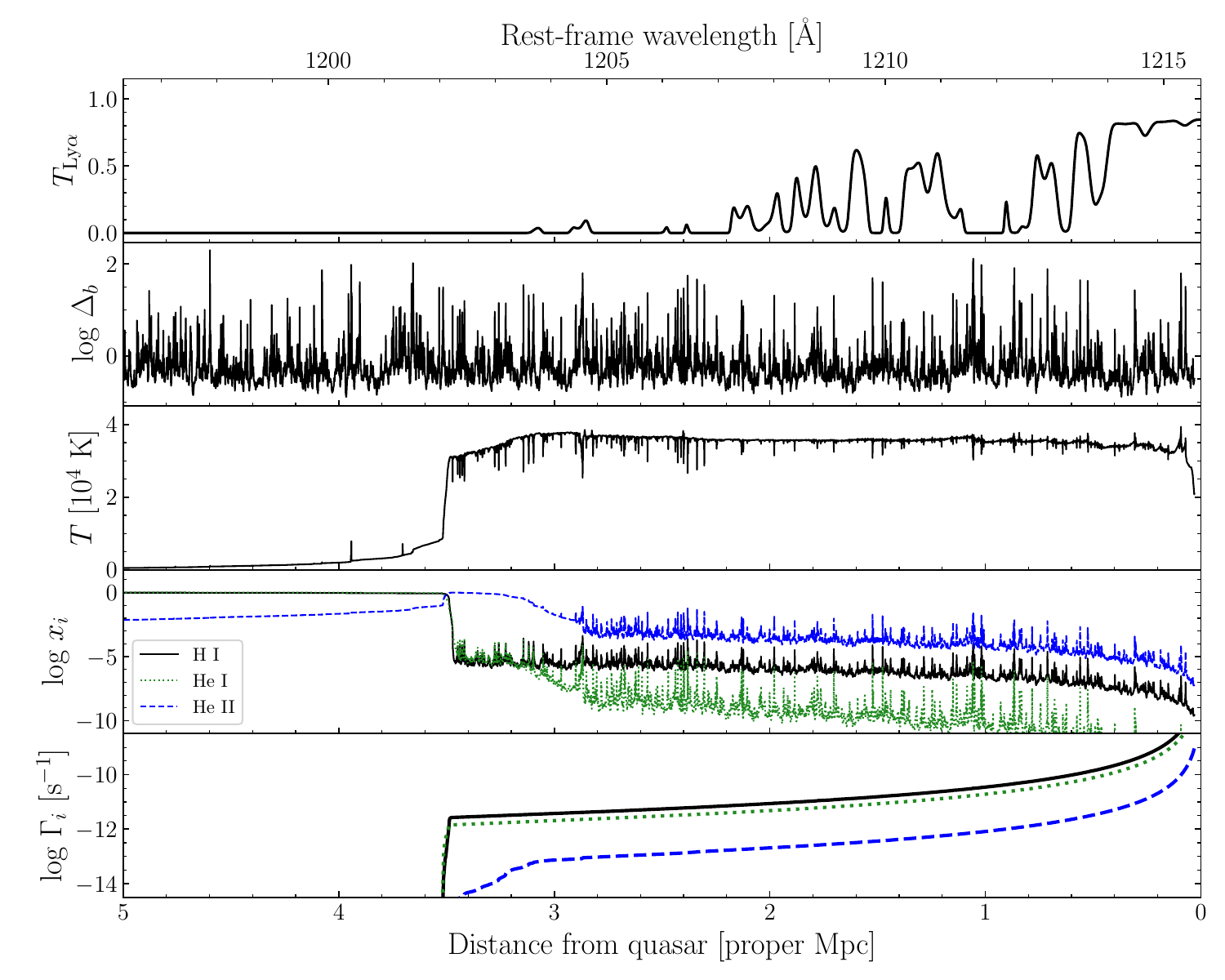}}
\end{center}
\caption{Example of a 1D radiative transfer simulation from the Cold CDM model. The panels from top to bottom show the Ly$\alpha$ transmission spectrum, the baryon overdensity, the gas temperature, the fractions of H\,{\small I}/\ion{He}{1}/He\,{\small II}, and the photoionization rates of H\,{\small I}/\ion{He}{1}/He\,{\small II}. }
\label{fig:rt_ex}
\end{figure*}

To simulate quasar proximity zones, we post-process skewers from the hydrodynamical simulations with the 1D hydrogen and helium ionizing radiative transfer (RT) code from \citet{Davies16}, originally based on \citet{BH07}, with additional minor modifications as described in \citet{Davies18b,Davies20a}. As mentioned above, the sizes of the ionized (and Ly$\alpha$-transparent) regions around $z\gtrsim7$ quasars are observed to be on the order of $\sim1$--$2$ proper Mpc. We must then stitch together many sightlines through our 1\,Mpc$/h$ hydrodynamical simulation box in order to fully capture the proximity zone morphology. In practice, we stitch together 58 random (without replacement) 1\,Mpc$/h$ (comoving) skewers to construct 10\,proper Mpc lines of sight. Each 1\,Mpc$/h$ skewer consists of 2048 evenly spaced samples\footnote{We oversample relative to the mean interparticle spacing to capture the smaller-scale density peaks that contribute substantially to the Ly$\alpha$ opacity.} of density, temperature, and velocity computed as a weighted sum of the SPH kernels of particles within a smoothing length of the skewer. For simplicity, we assume that the IGM is fully neutral at the start of each simulation\footnote{In principle, the X-ray models should be initially partially photoionized. However this initial ionization is very small $\lesssim1\%$ and thus neglecting it does not affect our results.} with no external ionizing background radiation. We assume a quasar UV absolute magnitude of $M_{1450}=-27$, convert to the specific luminosity at the Lyman limit using the \citet{Lusso15} spectral energy distribution, and then approximate the ionizing spectrum as a power law $L_\nu \propto \nu^{-1.7}$.

For each hydrodynamical simulation, we ran ionizing RT simulations through 1000 of the $10$ proper Mpc lines of sight with outputs in time logarithmically spaced from $10^4$ to $10^{7.25}$ yr in steps of 0.25 dex. For quasar activity substantially longer than $\sim10^7$ years, the hydrodynamical response of the gas to the rapid photoionization heating can no longer be ignored (e.g. \citealt{Shapiro04,D'Aloisio20}), but the proximity zone spectra of known $z>7$ quasars are more compatible with $\sim10^6$ year lifetimes \citep{Davies19}. We have also run a test simulation in which the IGM was reionized 10$^7$ years prior to $z=7.5$, and found that the baryonic structure at $z=7.5$ was negligibly affected. The long hydrodynamical response timescale implies that hints of the pre-reionization IGM structure may continue to be visible in spectra of quasars at lower redshifts into the end stages of reionization (e.g. \citealt{Doughty23}), but we leave an exploration of the long timescale decay to future work.

In Figure~\ref{fig:rt_ex} we show the physical properties of a simulated line of sight from the Cold CDM model after the quasar has been on for $10^7$\,yr. Compared to the relatively low resolution (but much larger volume) simulation skewer shown in Figure~1 of \citet{Davies20a}, the IGM has considerably more small-scale structure in the density field, and because the IGM was neutral beforehand, substantial heat has been injected by the ionization of hydrogen and helium leading to gas temperatures of roughly $40,000$\,K (see also \citealt{Davies16,Davies18b}). 

\subsection{Effect of Small-Scale Structure on Quasar Proximity Zones}

\begin{figure}
\begin{center}
\resizebox{8cm}{!}{\includegraphics[trim={1.0em 1em 1.0em 1em},clip]{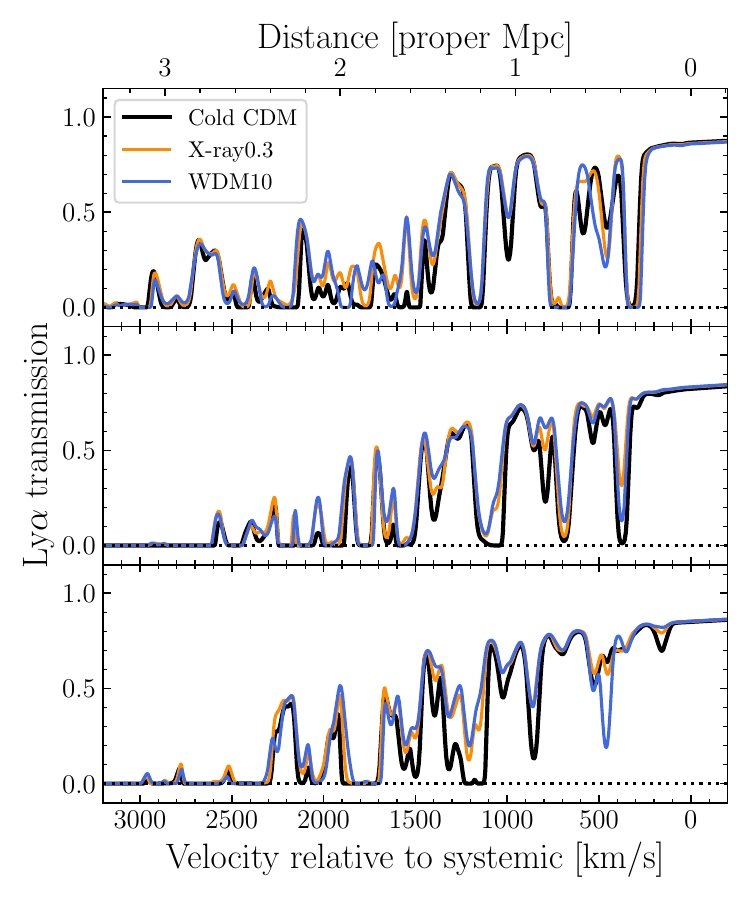}}
\end{center}
\caption{Comparison of Ly$\alpha$ transmission spectra from the Cold CDM (black), Xray0.3 (orange), and WDM10 (blue) models at the native 3\,km/s resolution of the RT simulation outputs. The three panels correspond to different lines of sight.}
\label{fig:rt_skew_vs_cdm}
\end{figure}

\begin{figure}
\begin{center}
\resizebox{8cm}{!}{\includegraphics[trim={4.0em 2.9em 1.5em 1.5em},clip]{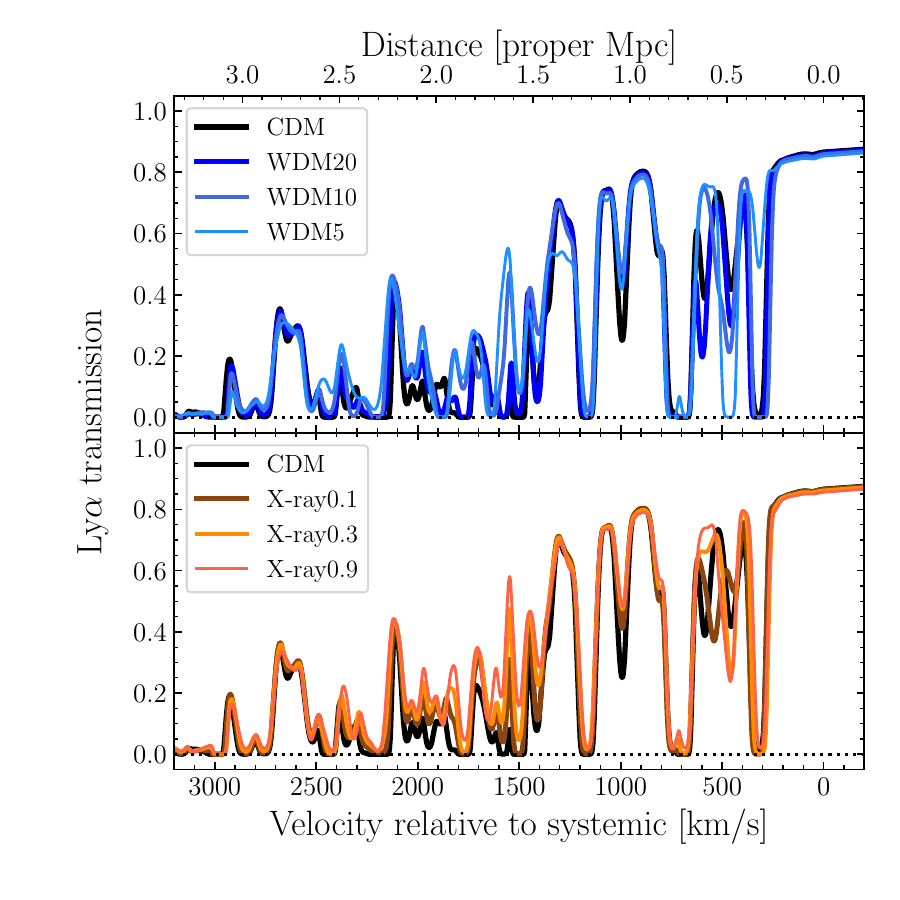}}
\end{center}
\caption{Comparison of Ly$\alpha$ transmission spectra between the Cold CDM model and the X-ray heated (top) and WDM (bottom) models. The spectra have been convolved to a spectral resolution $R=10000$ and binned to a pixel scale of $\Delta v=15$\,km/s, comparable to VLT/X-Shooter spectra.}
\label{fig:rt_skew_compare}
\end{figure}

In Figure~\ref{fig:rt_skew_vs_cdm}, we show the Ly$\alpha$ transmission spectra of simulated proximity zones in the Cold CDM, Xray0.3, and WDM10 simulations. Echoing the qualitative difference between the baryonic structures seen in Figure~\ref{fig:slices}, the Cold CDM Ly$\alpha$ spectra show an increased degree of structure, mostly notably deeper and sharper absorption features. Despite the $\sim40,000$\,K heat injection by the quasar as it ionized the hydrogen and helium in its surrounding IGM, blurring any small-scale structure in velocity space, the imprint of the dense, clumpy baryon structures is still visible due to the quadratic dependence of the Ly$\alpha$ opacity on gas density (e.g. \citealt{Weinberg97}).

In Figure~\ref{fig:rt_skew_compare}, we show a broader comparison between the Cold CDM, X-ray heated, and WDM models, for a single skewer from our simulation suite. In the top row we show the variations between the WDM models, while in the bottom row we show the different X-ray heated models, and the Cold CDM model is shown as the black curve in both panels. For this comparison, we smooth the spectrum with a Gaussian line spread function corresponding to $R=10000$ and re-bin to 15\,km/s pixels to emulate realistic ground-based observations with e.g. VLT/X-Shooter. We note that given the high gas temperatures reached inside the proximity zone ($\sim40000$\,K, cf.~Figure~\ref{fig:rt_ex}), this moderate spectral resolution is sufficient to nearly fully resolve the Doppler broadening of the Ly$\alpha$ transmission. In both sets of simulations we see a smooth trend of the spectral structure changing along with the smoothness of the physical structure. These observable differences in proximity zone structure between the models can be quantified by various Ly$\alpha$ forest summary statistics. 

\begin{figure}
\begin{center}
\resizebox{8cm}{!}{\includegraphics[trim={1.0em 1em 1.0em 1em},clip]{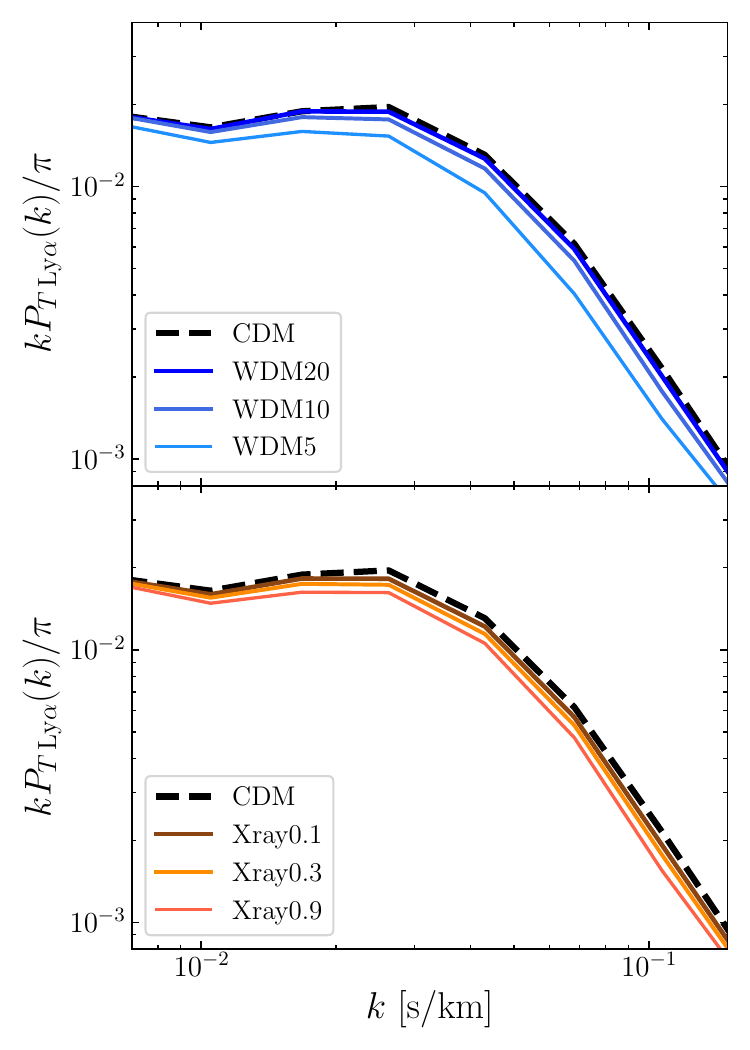}}
\end{center}
\caption{Dimensionless power spectra of $T_{{\rm Ly}\alpha}$ of the Cold CDM model (black) compared to the WDM (top) and Xray-heated (bottom) models for a quasar lifetime of 10$^7$ years.}
\label{fig:pk}
\end{figure}

\begin{figure}
\begin{center}
\resizebox{8cm}{!}{\includegraphics[trim={0.0em 1em 1.0em 1em},clip]{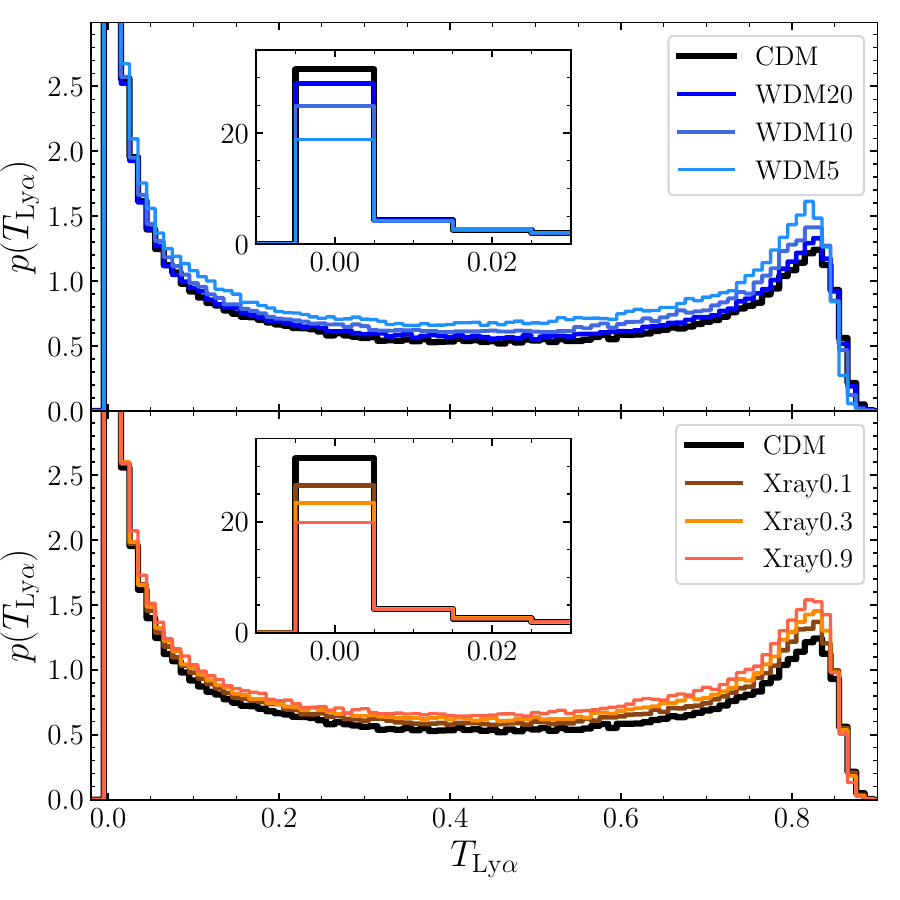}}
\end{center}
\caption{Probability distribution function of $T_{{\rm Ly}\alpha}$ from the Cold CDM model (black) compared to the WDM (top) and Xray-heated (bottom) models for a quasar lifetime of 10$^7$ years. The inset panels highlight the peak of saturated pixels at $T_{{\rm Ly}\alpha}\sim0$.}
\label{fig:pdf}
\end{figure}

First, we examine the flux power spectrum, $P_{\rm F}(k)$, which has previously been used to measure the small-scale structure of the Ly$\alpha$ forest (e.g. \citealt{Walther18,Walther19,Boera19,Gaikwad21}). The most common form of the flux power spectrum measures the power of $\delta_F\equiv(F-\bar{F})/\bar{F}$, where $\bar{F}$ is the mean transmitted flux. Because our $\bar{F}$ varies with position in the spectrum, we instead compute the power of $F$ directly. We restrict our analysis of the simulation outputs to 0 to 3000\,km/s from the quasar systemic frame, beyond which the flux is universally very close to zero, and average the modes into logarithmically binned band powers with dlog$k=0.2$. In Figure~\ref{fig:pk} we show the binned, dimensionless power spectra of $T_{{\rm Ly}\alpha}$ for the WDM and Xray-heated simulations, compared to Cold CDM, for a quasar lifetime of 10$^7$ years. For simplicity, we have not yet introduced noise and instrumental resolution effects -- these will be addressed in the next section.

Another often-used statistic of the Ly$\alpha$ forest is the transmitted flux PDF (e.g. \citealt{Lidz06,Bolton08,Rollinde13,Lee15}). We note from Figure~\ref{fig:rt_skew_vs_cdm} that the Cold CDM models appear to have more saturated pixels than the models with suppression of small-scale structure. We thus expect a corresponding signal to be apparent in the PDF of $T_{{\rm Ly}\alpha}$. In Figure~\ref{fig:pdf}, similar to Figure~\ref{fig:pk}, we show the flux PDFs from our simulations in bins of $\Delta F=0.01$ assuming $t_{\rm Q}=10^7$\,yr, with inset panels to highlight the difference in the number of saturated pixels with $T_{\rm Ly\alpha}\sim0$. As expected, the simulations with more small-scale structure show more saturated pixels. 

\section{Constraining Power of Quasar Proximity Zones}

The differences in proximity zone structure between the Cold CDM model and models with suppressed small-scale structure shown in Figure~\ref{fig:rt_skew_vs_cdm} are visible by eye, and from Figures~\ref{fig:pk} and \ref{fig:pdf} there are apparent differences in Ly$\alpha$ forest statistics, however it remains to be seen whether these differences can be detected statistically given the substantial cosmic variance of the IGM along any given line of sight, and the finite signal-to-noise achievable in realistic observations. Here we explore the constraining power of the flux power spectrum and the flux PDF, as well as the combination of the two. Given the sparsity of our simulation suite, we adopt a model comparison strategy rather than a parameter inference one, to assess the degree to which the models with suppressed small-scale structure can be distinguished from the fiducial Cold CDM model.

For the tests below, we first process the simulated proximity zone spectra into mock observations in a VLT/X-Shooter-like configuration by re-binning the spectra to 15\,km/s pixels and convolving to a Gaussian spectral resolution of $R=10000$. We further assume a signal-to-noise ratio ${\rm S/N}=30$ per pixel of the unabsorbed continuum, comparable to existing deep X-Shooter spectra of $z\gtrsim7$ quasars (e.g.~\citealt{Bosman17,D'Odorico23}). We then compute the flux power spectrum and flux PDF statistics described in the previous section using pixels within 3000\,km/s of the systemic redshift of the quasar, roughly corresponding to the maximum distance of pixels with detectable Ly$\alpha$ transmission in the simulated proximity zone spectra.

We then perform two tests of the constraining power of each summary statistic, first comparing Cold CDM to the WDM models, and then comparing Cold CDM to the Xray-heated models. In both cases, we have simulated spectrum outputs for the Cold CDM model plus three additional models (WDM5/WDM10/WDM20 or Xray0.1/Xray0.3/Xray0.9), which we denote by $\mathcal{M}$, and 14 quasar lifetimes $t_{\rm Q}$ from $10^4$ to $10^{7.25}$\,yr for each model. We adopt a statistical methodology similar to \citet{Davies18b} -- in brief, we construct an approximate likelihood from the simulated spectra, optimize that likelihood for a given mock data set, use the maximum (approximate) likelihood model parameters (on our fixed grid of models) as the summary statistic, and then directly compute the likelihood function of the summary statistic via sampling a large number of mock data sets. 

We start by constructing 20,000 mock sets of one, four and ten $z=7.5$ quasar spectra by drawing without replacement from the set of 1,000 RT spectra and adding a spectral noise realization. We then compute the mean $\mu$ and covariance matrix $\Sigma$ of the summary statistics (i.e. $P_{\rm F}(k)$ and/or $p(F)$) from the mock ensembles of quasar spectra for each model $\theta=\{\mathcal{M},t_{\rm Q}\}$, and approximate the likelihood function, $\tilde{L}(\theta)$, as a multivariate Gaussian $\mathcal{N}(\mu(\theta),\Sigma(\theta))$. For each mock data set, we evaluate $\tilde{L}$ for each of the $4\times14=56$ models, and record the model parameters $\theta_{\rm M\tilde{L}E}$ for which $\tilde{L}$ is maximized. We perform this procedure for each true model $\theta$, and build up a mapping between $\theta$ and $\theta_{\rm M\tilde{L}E}$. From an individual mock data set, resulting in an individual measurement of $\theta_{\rm M\tilde{L}E}$, we can then compute the posterior PDF of $\theta$, $p(\theta|\theta_{\rm M\tilde{L}E})$, via Bayes' theorem,
\begin{equation}
    p(\theta|\theta_{\rm M\tilde{L}E}) = \frac{p(\theta_{\rm M\tilde{L}E}|\theta)p(\theta)}{p(\theta_{\rm M\tilde{L}E})},
\end{equation}
where $p(\theta_{\rm M\tilde{L}E}|\theta)$ is the likelihood function measured directly from the Monte Carlo mock observations and $p(\theta)$ is an assumed prior on the model parameters $\theta$. The evidence $p(\theta_{\rm M\tilde{L}E})$, is also computed directly from the Monte Carlo mocks,
\begin{equation}
    p(\theta_{\rm M\tilde{L}E}) = \int p(\theta_{\rm M\tilde{L}E}|\theta) p(\theta) d\theta
\end{equation}
For simplicity, we assume a uniform prior on $\mathcal{M}$ and $\log{t_{\rm Q}}$.

When considering a data set with more than one quasar, there are multiple possible ways to construct a joint posterior PDF on $\mathcal{M}$. One way is to maximize the total $\tilde{L}$ from all quasars assuming a single value for the quasar lifetime. We will adopt this method when showing joint 2D posterior PDFs for the sake of illustration. The more conservative approach, which we adopt for our final prospective analysis, is to compute the individual 2D posterior PDFs for each quasar, marginalize out the unknown quasar lifetime, and then combine the resulting 1D posteriors into a final posterior PDF on $\mathcal{M}$. 

\subsection{Flux power spectrum}

\begin{figure}
\begin{center}
\resizebox{8.5cm}{!}{\includegraphics[trim={0.0em 1em 1.0em 1em},clip]{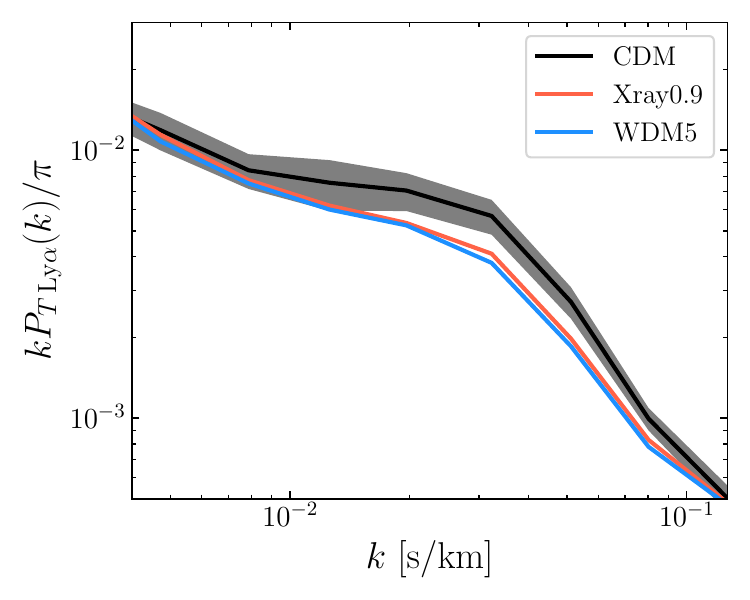}}
\end{center}
\caption{Forward-modeled dimensionless power spectra of $T_{{\rm Ly}\alpha}$ of the Cold CDM model (black) compared to the WDM (top) and Xray-heated (bottom) models for a quasar lifetime of 10$^{6.5}$ years. The shaded region shows the $1\sigma$ scatter of mock observations consisting of 10 quasars each.}
\label{fig:pk2}
\end{figure}

We first apply the method described above to the flux power spectrum. We restrict our analysis to $k<0.1$\,s/km, as the band powers on smaller scales are dominated by noise and thus do not contribute any additional statistical discriminating power. In Figure~\ref{fig:pk2}, we show flux power spectra from Cold CDM and the two most distant models, WDM5 and Xray0.9, with forward-modeled noise, binning, and spectral resolution. The shaded region shows the $1\sigma$ scatter of the PDF measurements in each bin assuming the Cold CDM model as the truth and a sample size of 10 quasars with lifetimes of $10^{6.5}$\,yr.

\begin{figure*}
\begin{center}
\resizebox{8.5cm}{!}{\includegraphics[trim={0.0em 1em 1.0em 1em},clip]{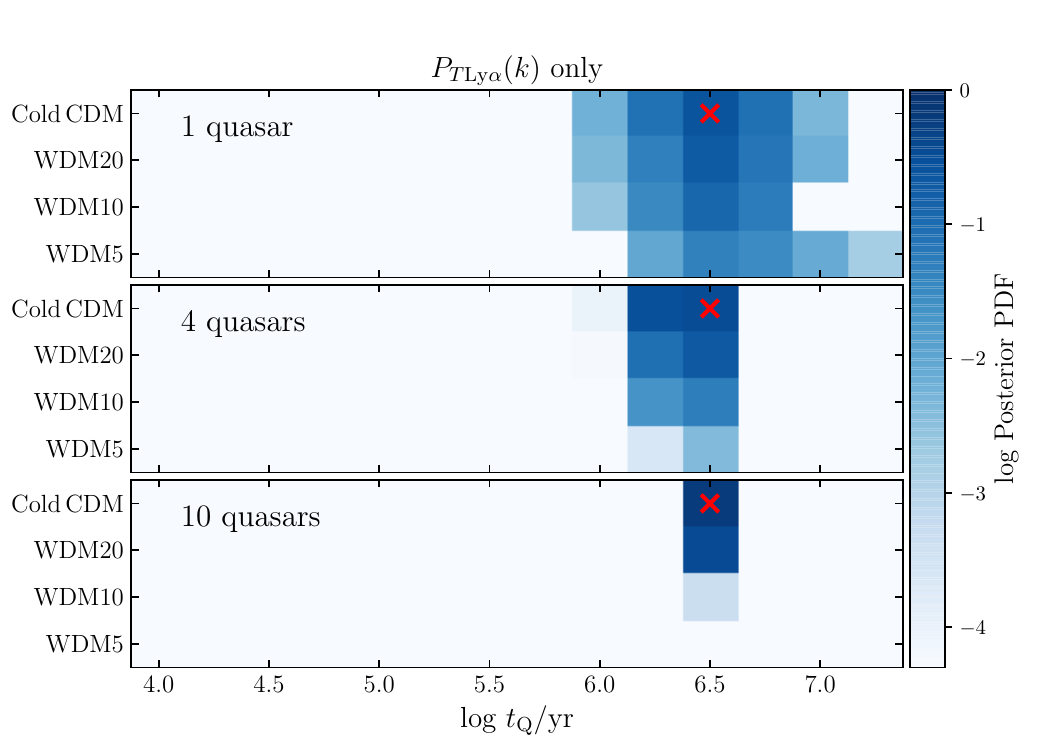}}
\resizebox{8.5cm}{!}{\includegraphics[trim={0.0em 1em 1.0em 1em},clip]{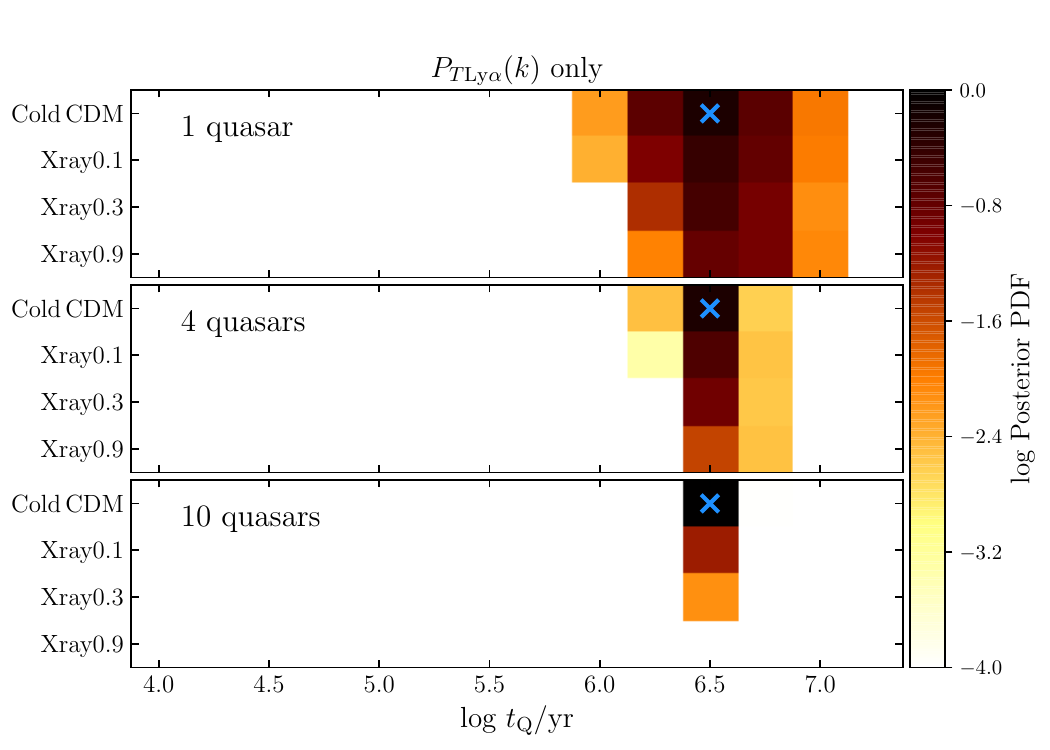}}
\end{center}
\caption{Two-dimensional posterior PDFs on $\{\mathcal{M},t_{\rm Q}\}$ for the WDM models (left) and Xray-heated models (right) compared to Cold CDM. From top to bottom, the panels show example posterior PDFs for samples of 1, 4, and 10 quasars, where the cross marks the model assumed to be the ``true'' model.}
\label{fig:pk_2d_post}
\end{figure*}

In Figure~\ref{fig:pk_2d_post}, we show examples of 2D posterior PDFs of $\{\mathcal{M},t_{\rm Q}\}$ assuming as truth the Cold CDM model with true $t_{\rm Q}=10^{6.5}$\,yr -- the same lifetime for every quasar -- versus the WDM models and versus the Xray-heated models. We show examples for data sets of one, four, and ten $z=7.5$ quasars from top to bottom, where we combine the posterior PDFs from individual quasars by multiplying the 2D posteriors directly -- i.e., for illustration, we assume that all quasars have the same lifetime. While one quasar does not have much constraining power in the $\mathcal{M}$ dimension, samples of four and ten quasars may be capable of distinguishing between the Cold CDM model and some range of WDM or X-ray-heated models.

\subsection{Flux PDF}

\begin{figure}
\begin{center}
\resizebox{8.5cm}{!}{\includegraphics[trim={0.0em 1em 1.0em 1em},clip]{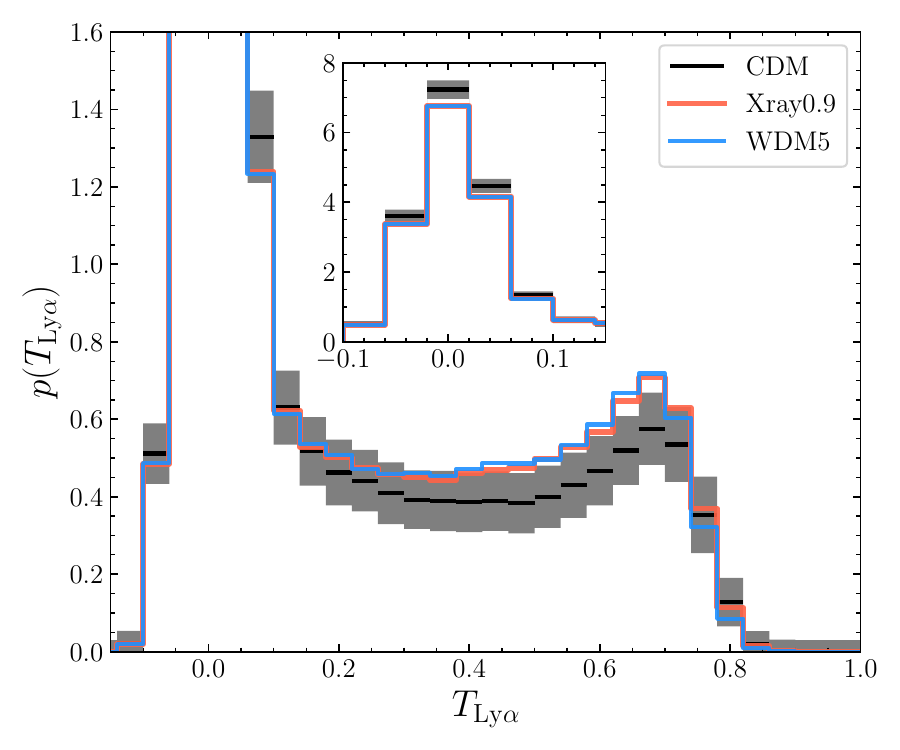}}
\end{center}
\caption{Similar to Figure~\ref{fig:pk2} but for the forward-modeled transmitted flux PDF. The inset panel highlights the behavior of the PDF close to a transmitted flux of zero.}
\label{fig:pdf2}
\end{figure}

\begin{figure*}
\begin{center}
\resizebox{8.5cm}{!}{\includegraphics[trim={0.0em 1em 1.0em 1em},clip]{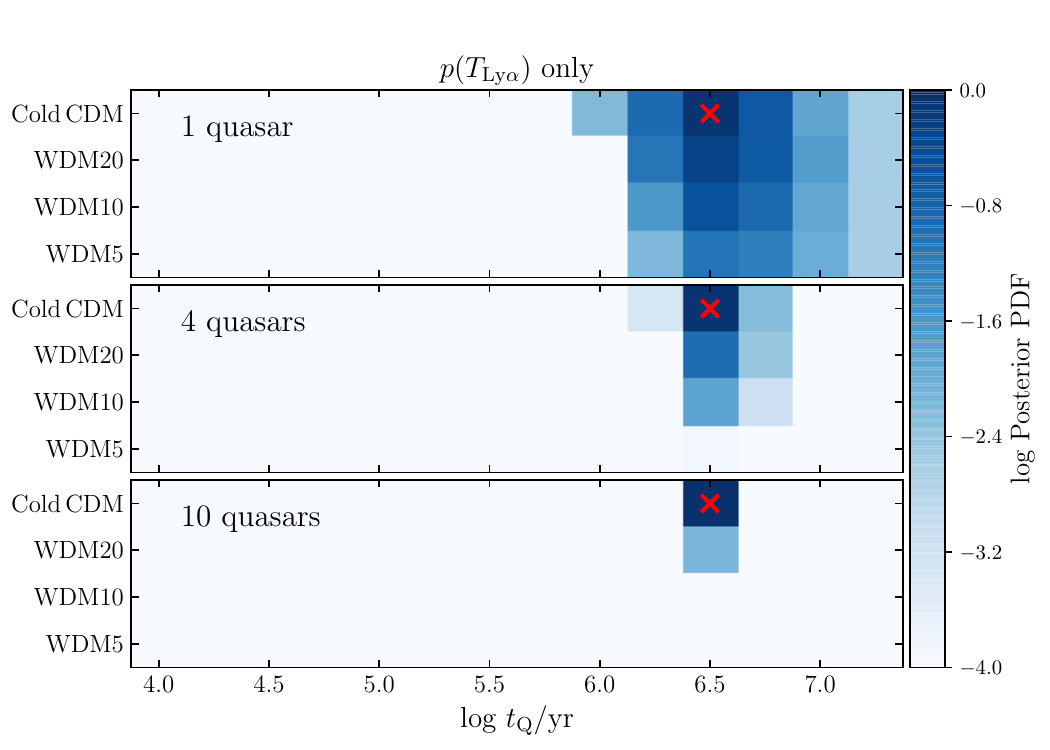}}
\resizebox{8.5cm}{!}{\includegraphics[trim={0.0em 1em 1.0em 1em},clip]{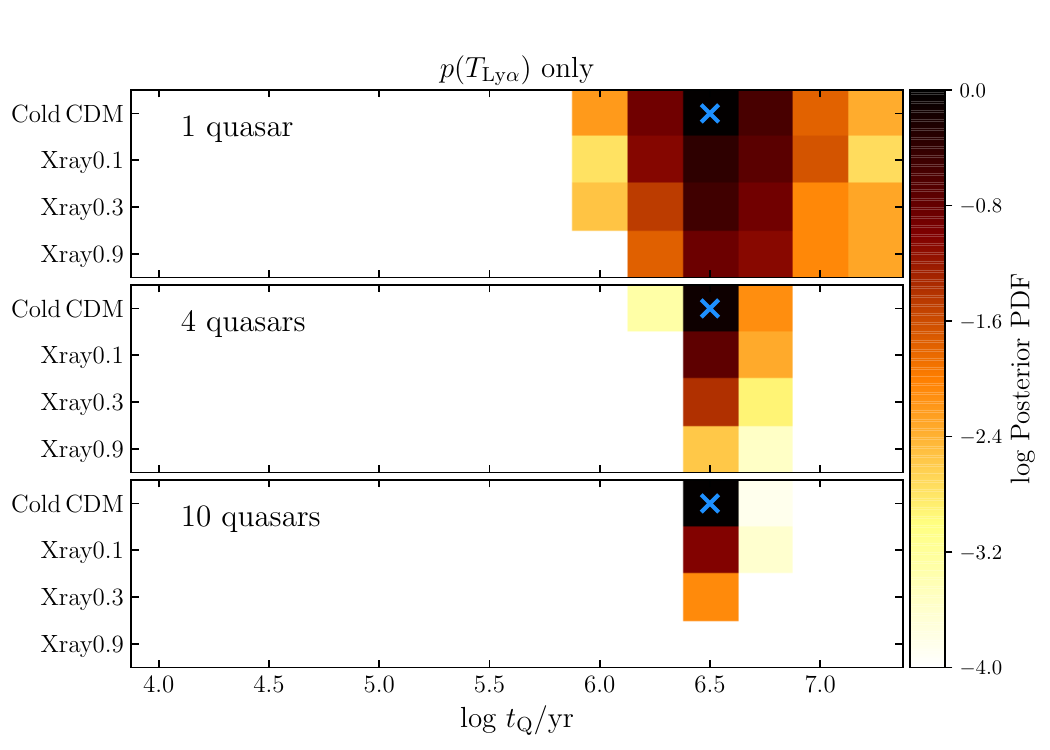}}
\end{center}
\caption{Similar to Figure~\ref{fig:pk_2d_post} but for the transmitted flux PDF.}
\label{fig:pdf_2d_post}
\end{figure*}

To estimate the constraining power of the flux PDF, we employ the same statistical analysis technique as described in the previous section using bins of $\Delta T_{{\rm Ly}\alpha}=0.04$ from $T_{{\rm Ly}\alpha}=-0.16$--$1.08$. The forward-modeled flux PDFs for the Cold CDM, WDM5, and Xray0.9 models are shown in Figure~\ref{fig:pdf2}. Because some PDF bins are not populated in some regions of the parameter space, we regularize the covariance matrix by adding a small constant\footnote{We choose this constant to be the equivalent Poisson variance of an average occurrence rate of one per quasar sample.} along the diagonal, finding that the exact value of this constant does not affect our results. In Figure~\ref{fig:pdf_2d_post} we show the resulting 2D posterior PDFs, analogous to Figure~\ref{fig:pk_2d_post}. In general we find that the flux PDF is more constraining than the power spectrum. 

\subsection{Joint constraints}

\begin{figure*}
\begin{center}
\resizebox{8.5cm}{!}{\includegraphics[trim={0.0em 1em 1.0em 1em},clip]{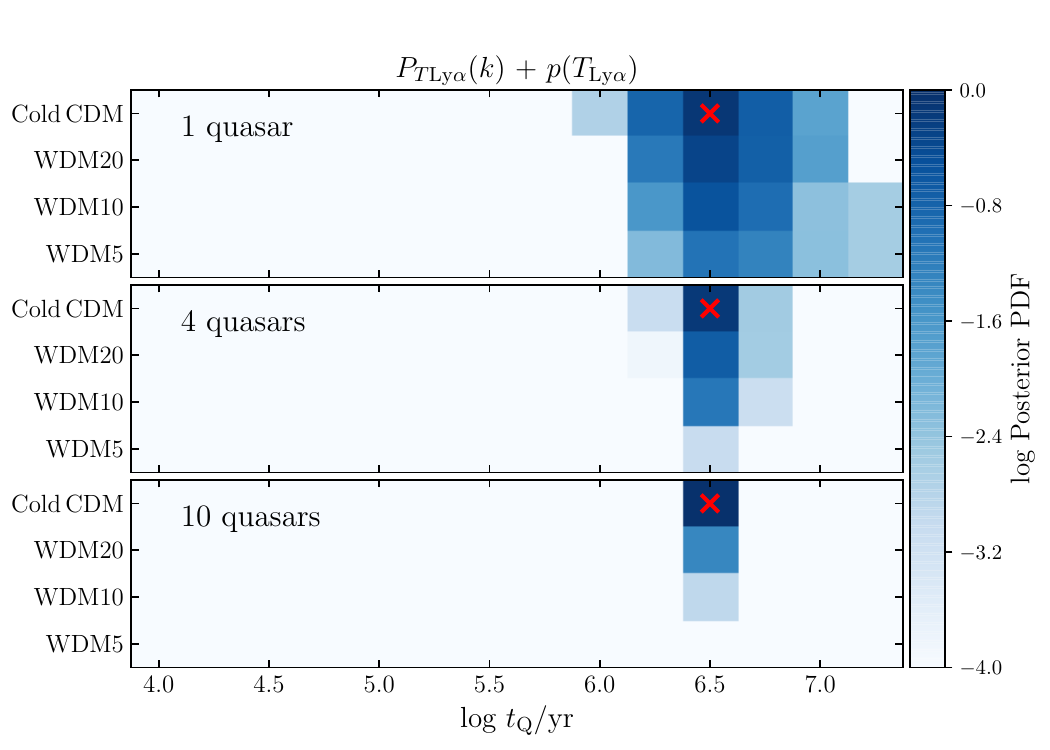}}
\resizebox{8.5cm}{!}{\includegraphics[trim={0.0em 1em 1.0em 1em},clip]{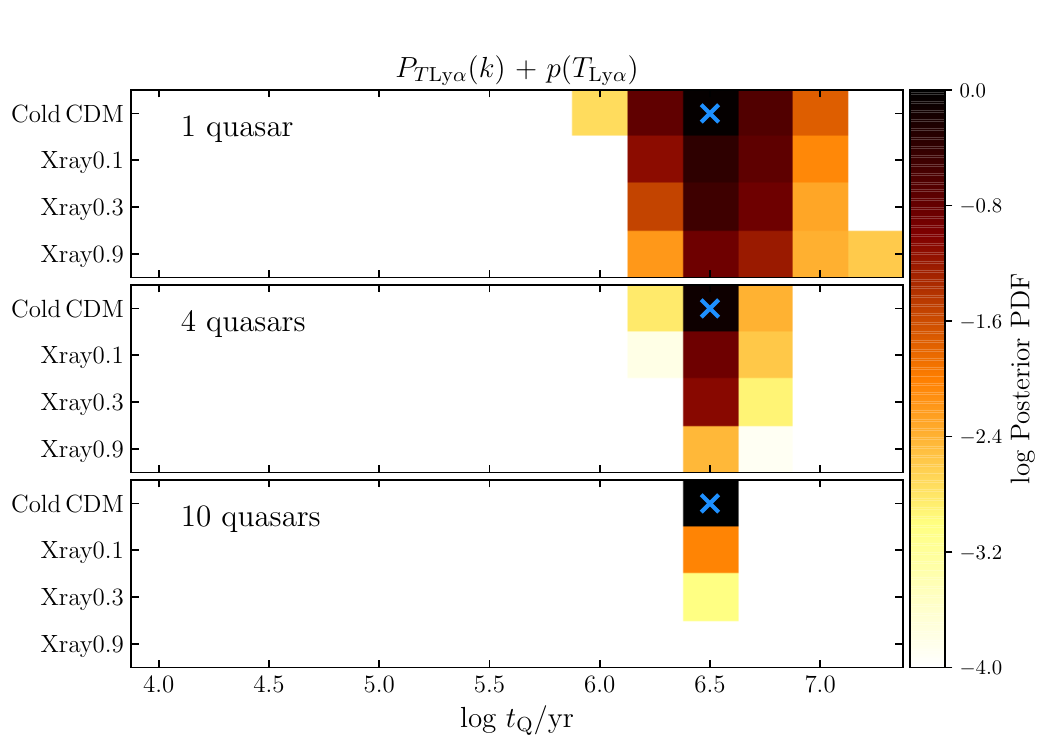}}
\end{center}
\caption{Similar to Figures~\ref{fig:pk_2d_post} and \ref{fig:pdf_2d_post} but for joint constraints from the power spectrum and flux PDF.}
\label{fig:both_2d_post}
\end{figure*}

Finally, our statistical analysis method allows for a straightforward combination of the constraints from the two summary statistics without a computationally challenging assessment of their cross-covariance (e.g. \citealt{Wolfson21}). We do this by defining the pseudo-likelihood as the product of the pseudo-likelihoods for $P_{T{\rm Ly}\alpha}(k)$ and $p(T_{{\rm Ly}\alpha})$. Because we ignore the likely considerable covariance between the two statistics for simplicity, our pseudo-likelihood is not entirely optimal. However, the calibration procedure of mapping the maximum pseudo-likelihood parameters to the true ones ensures that the resulting posterior PDF is nevertheless unbiased. The resulting 2D posterior distributions vs. WDM and X-ray heated models are shown in Figure~\ref{fig:both_2d_post}. Combining the less constraining power spectrum with the more constraining flux PDF, with equal pseudo-likelihood weights, is slightly more discriminating than the flux PDF alone.

\subsection{Model comparison}\label{sec:modelcompare}

We perform the analysis described above on 20,000 mock data sets generated from 1,000 forward-modeled quasar spectra to investigate the range of possible posterior PDFs in the ``model'' dimension $\mathcal{M}$, i.e. Cold CDM vs. WDM or Xray-heated models. As our model grid is coarse, rather than interpolating or treating the posterior PDF as a quantitative constraint on the particular model parameter (i.e. WDM mass or temperature of X-ray heating) we instead consider the statistical significance with which the Cold CDM model can be distinguished from the other models. Towards this end, we investigate the ratio between the likelihood function evaluated at the Cold CDM model to the other models. 

\begin{figure}
\begin{center}
\resizebox{8.5cm}{!}{\includegraphics[trim={0.0em 1em 1.0em 1em},clip]{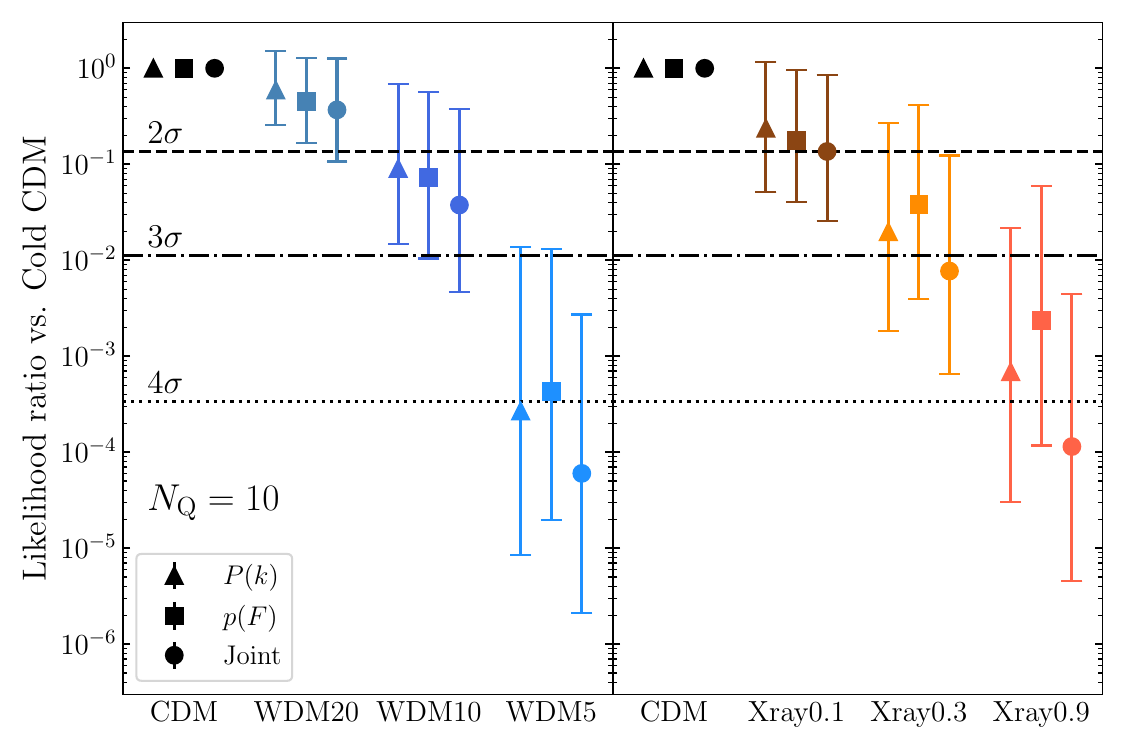}}
\end{center}
\caption{Likelihood ratio between the Cold CDM model and WDM (left) and Xray-heated (right) models, for mock data sets of ten quasars drawn from the Cold CDM model. Medians (points) and central 68\% distributions (error bars) are shown for constraints using the flux power spectrum (triangles), flux PDF (squares), and joint constraints from a combination of the two (circles). The dashed, dot-dashed, and dotted lines show likelihood ratios corresponding to $2\sigma$, $3\sigma$, and $4\sigma$, respectively.}
\label{fig:ratio}
\end{figure}

Note that in the 2D posteriors from four and ten quasars shown in Figures~\ref{fig:pk2}, we have implicitly assumed that all quasars have the same lifetime. For this model comparison test, we still assume that each quasar in the simulated samples has the same true lifetime of $10^{6.5}$\,yr, but now marginalize over the individual lifetime of each quasar before multiplying the 1D likelihood functions in the model dimension, and thus our final combined likelihood does not assume that all quasars have the same lifetime.

In Figure~\ref{fig:ratio}, we show the median and central 68\% scatter of the likelihood ratio between the Cold CDM model and the WDM (left) and Xray-heated models (right) for mock observed samples of ten quasars. Low values of the likelihood ratio indicate strong discriminating power of the data between the corresponding model and the Cold CDM model. We find that the flux PDF statistic is substantially more constraining than the flux power spectrum, and that combining the two statistics (at least in our conservative statistical approach) modestly improves the total constraining power. The horizontal lines in Figure~\ref{fig:ratio} represent $2\sigma$, $3\sigma$, and $4\sigma$ equivalent discrimination versus the Cold CDM model from top to bottom. 

Due to the small volume probed by even a sample of ten quasars, cosmic variance in the IGM structures probed by the particular quasars causes the the achieved constraining power to vary considerably between the mock data sets. That is, there is considerable scatter in the likelihood ratio between different IGM models and Cold CDM between different mock data sets. At the upper $1\sigma$ end of the constraining power (i.e. the lower error bars in Figure~\ref{fig:ratio}), even the WDM20/Xray0.1 models can be distinguished from Cold CDM at almost $3\sigma$, while at the lower $1\sigma$ end (i.e. the upper error bars in Figure~\ref{fig:ratio}), the WDM10/Xray0.3 models would only be distinguished at the $\lesssim2\sigma$ level. Inspecting the median ratios shown by the points in Figure~\ref{fig:ratio}, which we treat as fiducial, we see that the joint statistic is typically incapable of distinguishing between the Cold CDM and WDM20 models, but the Xray0.1 model may be distinguishable at $\sim2\sigma$, the WDM10/Xray0.3 models can be distinguished at $\sim3\sigma$, and the WDM5/Xray0.9 models at the $>4\sigma$ level.

\section{Discussion \& Conclusion}

In this work, we showed that the small-scale structure of the pre-reionization IGM may have a detectable impact on the proximity zones of the highest redshift quasars. We first ran a suite of small-volume hydrodynamical simulations using \texttt{MP-Gadget} with varying WDM particles masses and heat injection by X-rays down to $z=7.5$. We customized the simulations for our particular implementation by carefully constructing initial conditions accounting for the baryon-dark matter streaming velocity and imposing a floor to the electron fraction due to freeze-out following recombination. We then performed 1D radiative transfer simulations of quasar ionizing radiation through skewers from the simulation suite, and computed the flux power spectrum and flux PDF statistics from the resulting mock Ly$\alpha$ proximity zone spectra. Finally, we showed that measurements of these statistics from future samples of $z>7$ quasars should enable compelling constraints on the properties of dark matter and early X-ray heating, with typical $2\sigma$ limits of $m_{\rm WDM}\gtrsim10$--$20$\,keV and $f_{\rm X}f_{\rm abs}\lesssim0.1$ ($T_{\rm IGM}\lesssim275$\,K). These constraints depend crucially on a handful of major assumptions which we discuss below.

First, we have assumed that the IGM along the line of sight to the quasars is entirely neutral when they turn on. That is, the surrounding IGM has not been ``pre-reionized'' by other sources, e.g. galaxies. Such pre-reionized regions are theoretically expected from the clustering of galaxies around the massive dark matter halos thought to host quasars (e.g. \citealt{AA07}). However, current observations $z\gtrsim7$ quasars are consistent with a fully neutral initial state, unless one invokes extremely short quasar lifetimes $t_{\rm Q} \ll 10^6$\,yr \citep{Davies19}. Even if pre-reionized gas is present, one can limit its impact by observing quasars at yet higher redshifts. Finally, as shown in \citet{Davies18b}, it is possible to include the impact of pre-reionized regions in the model, i.e. we could include the IGM neutral fraction $\langle x_{\rm HI} \rangle$ as an additional parameter. We opt for a smaller parameter space here for simplicity, but similar to $t_{\rm Q}$ we do not expect significantly weaker constraining power if we were to include $\langle x_{\rm HI}\rangle$. A perhaps more worrisome issue is if the regions were pre-reionized long enough prior to the quasar turning on that the gas has time to hydrodynamically respond to the reionization heating; e.g., there would have been $\sim100$\,Myr for the gas to respond if the region around the $z=7.5$ quasar was reionized at $z=8.5$. Reducing this uncertainty would require additional modeling of the gas response to a range of past times of reionization (e.g.~\citealt{Doughty23}). 

Second, we assume that the UV-luminous lifetime of the quasars is short enough that the gas has not had sufficient time to hydrodynamically respond to its new gas pressure. Observations of quasar proximity zones at $z>6$ suggest typical UV-luminous lifetimes of $10^6$\,yr \citep{Bosman20a,Eilers20,Eilers21,Morey21}, consistent with lifetime constraints from \ion{He}{2} proximity zones at lower redshift \citep{Khrykin19,Worseck21}, and in particular the highest redshift quasars at $z>7$ show strong indications of UV-luminous lifetimes no longer than $\sim10^6$\,yr \citep{Davies19}. Given the long relaxation time of the gas, $\gtrsim100$\,Myr \citep{Park16,D'Aloisio20}, the assumption of a static density field before and during the quasar's activity is likely a reasonable one.

Finally, we have assumed that the environment of the quasar is comparable to the average IGM. As mentioned above, luminous $z>7$ quasars are likely hosted by some of the most massive dark matter halos in existence at the time, and thus likely reside within larger Mpc-scale overdensities. Unfortunately, simulating large enough cosmological volumes ($>100$\,Mpc on a side) at high enough resolution ($\sim1$\,kpc) to both sample the high mass end of the halo mass function while simultaneously resolving the smallest structures in the IGM is currently computationally infeasible. Progress may still be possible via zoom-in simulations centered on massive halos (e.g. \citealt{Trebitsch20}), which we will explore in future work. We note that \citet{Davies20a} found that the effect of including a $>10^{11}$\,$M_\odot$ halo at the quasar position had only a minor effect on the typical proximity zone transmission profile, but their simulations did not include the small-scale structure studied here.

\begin{figure}
\begin{center}
\resizebox{8.5cm}{!}{\includegraphics[trim={0.0em 1em 1.0em 1em},clip]{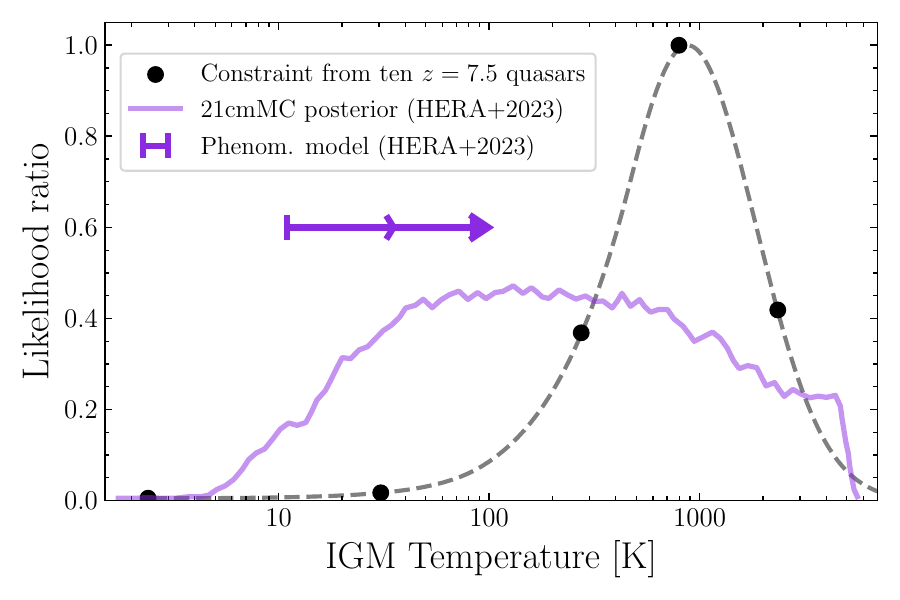}}
\end{center}
\caption{Median likelihood ratio for the X-ray heated models assuming the Xray0.3 model is the truth (black points), expressed in terms of their corresponding IGM temperature at $z=7.5$. An illustrative log-space quadratic spline interpolation is shown by the grey dashed curve. The purple curve shows the posterior PDF from \citet{HERA23} using 21cmMC \citep{GM15,GM17}, while the flat and angled segments of the purple arrow show the 95\% and 68\% limits, respectively, using the phenomenological approach of \citet{Mirocha22}.}
\label{fig:tpost}
\end{figure}

It is worth putting our sensitivity to X-ray heating in the context of 21 cm cosmology. The 21 cm signal is sensitive to the difference between the background CMB temperature $T_{\rm CMB}$ and the spin temperature of the 21 cm transition $T_{\rm S}$, $\delta T_{\rm b}\propto (T_{\rm S}-T_{\rm CMB})/T_{\rm S}$. Assuming that the spin temperature is fully coupled to the IGM kinetic temperature $T_{\rm IGM}$ via the Wouthuysen-Field effect \citep{Wouthuysen52,Field58}, which is almost certainly the case at the epoch we consider (e.g. \citealt{Meiksin23}), the 21 cm signal is strongest when the IGM is very cold but then saturates once $T_{\rm S} \gg T_{\rm CMB}$. That is, while non-detections of the 21 cm signal can already rule out an arbitrarily cold IGM at $z\sim8$--$10$ \citep{HERA22,HERA23}, at $z=7.5$ where $T_{\rm CMB}=23.2$\,K they are currently unable to distinguish between IGM temperatures of e.g. 100\,K and 1000\,K. Meanwhile, we find that the small-scale structure of the IGM probed by quasar proximity zones is more sensitive to higher temperatures $T_{\rm IGM}\sim1000$\,K. 

To explore the constraining power of our model in the context of an X-ray heated IGM, we redo the analysis from \S~\ref{sec:modelcompare} assuming that the true model is Xray0.3 ($T_{\rm IGM}=800$\,K), and include one additional model with very mild X-ray heating (Xray0.01) to fill in the $T_{\rm IGM}$ gap between Cold CDM and Xray0.1. In Figure~\ref{fig:tpost}, we show an illustrative likelihood ratio curve for $T_{\rm IGM}$, corresponding to a quadratic spline through the median likelihood ratios relative to the Xray0.3 model, that could be obtained from quasar proximity zones compared to the most recent constraints from \citep{HERA23}. Two of the HERA constraints on $T_{\rm IGM}$ are shown: lower limits from a bubble-driven phenomenological model of reionization \citep{Mirocha22}, and a posterior PDF from an astrophysically-motivated model using the 21cmFAST code \citep{Mesinger11,GM17,Park19}, wherein the constraining power at the upper end is driven by priors on the properties of high-redshift galaxies. Thus future proximity zone measurements from high-redshift quasar samples will be highly complementary to the lower temperature limit probed by 21-cm.

There is already possible tentative evidence for the existence of small-scale clumpy structures in the $z\gtrsim6$ IGM from the recent measurement of the ionizing photon mean free path by \citet[][see also \citealt{Zhu23,Gaikwad23,Davies23}]{Becker21}, thought to be largely driven by self-shielding of dense gas clouds (e.g.~\citealt{McQuinn11}). Reproducing the short observed value ($<1$\,proper Mpc) is only possible if the baryons trace (cold) dark matter down to small scales \citep{Emberson13}. Confirmation of small-scale structure in quasar proximity zones at $z>7$ would support the existence of a short ionizing photon mean free path throughout the reionization epoch, and thus place steep requirements on ionizing photon emission during the process \citep{Cain21,Davies21}.

Investigations of $z>7$ quasar damping wings and lifetimes using their proximity zones \citep{Davies18b,Davies19,Wang20,Yang20a} have thus far ignored the additional small scale structure described here, in the sense that the adopted hydrodynamical simulations are lower resolution by a factor of $>10$ in linear dimension. While it is difficult to assess this in detail, by comparing the average Ly$\alpha$ transmission profiles of the Cold CDM and Hot CDM models, where the latter is more representative of the simulations used in previous work, we find that the Ly$\alpha$-transparent proximity zones are slightly smaller on average in the Cold CDM case. However, the physical size of the ionized regions are almost identical, because the excess recombination rate from the dense peaks is still not high enough to significantly slow down the ionization front on timescales comparable to the quasar lifetime ($\sim10^6$-$10^7$\,yr). Thus we do not expect the damping wing signatures to change appreciably, except potentially through differences in the pre-existing ionized regions due to differences in the mean free path (e.g. \citealt{DF22}). 

Currently, only ten quasars are known at $z>7$, with only three known at $z>7.5$ (\citealt{Banados18,Yang20a,Wang21}; see also \citealt{BosmanList} and references therein). Searches for more $z>7$ quasars are still ongoing, with ever-improving methods of selecting candidates (e.g.~\citealt{Nanni22}). Future near-infrared surveys with \emph{Euclid} and \emph{Roman} will enable many more discoveries as well (e.g. \citealt{Barnett19}). Precision follow-up spectroscopy will be possible with JWST and the ELTs, and through methods similar to those described in this work, enable powerful constraints on pre-reionization IGM physics.

\bibliographystyle{aasjournal}
 \newcommand{\noop}[1]{}

\end{document}